\documentclass[superscriptaddress,nofootinbib,twocolumn,letterpaper,prx]{revtex4-1}

\usepackage{amsmath,amssymb}
\usepackage{graphicx}
\usepackage[usenames,dvipsnames]{xcolor}
\usepackage{hyperref}
\usepackage{url}
\usepackage[utf8]{inputenc}
\usepackage{slashed}
\usepackage{subfigure}
\usepackage{slashed,bbm}
\usepackage{graphicx,psfrag,epsfig}
\usepackage{dsfont}
\usepackage{setspace}
\usepackage{enumerate}
\usepackage{blindtext}
\usepackage{tikz}
\usetikzlibrary{arrows, calc, 3d, decorations.markings,decorations.pathmorphing,shapes,positioning,arrows.meta, fadings}
\hypersetup{
    colorlinks,
    linkcolor={red},
    citecolor={blue},
    urlcolor={blue!40!black}
}

\def\code#1{\texttt{#1}}

\newcommand*{\vcenteredhbox}[1]{\begingroup
\setbox0=\hbox{#1}\parbox{\wd0}{\box0}\endgroup}
\newcommand{\picscalefactor}{0.5}

\allowdisplaybreaks

\begin{document}

\title{Abelian Higgs model at four loops, fixed-point collision and deconfined criticality}

\begin{flushright}
      \normalsize HU-EP-19/19, DESY 19-124, SAGEX-19-16
\end{flushright}
\vspace{-0.8cm}

\author{Bernhard Ihrig}
\affiliation{Institute for Theoretical Physics, University of Cologne, Z\"ulpicher Stra\ss e 77, Cologne, Germany}

\author{Nikolai Zerf}
\affiliation{Institut f\"ur Physik, Humboldt-Universit\"at zu Berlin, Newtonstraße 15, Berlin, Germany}

\author{Peter Marquard}
\affiliation{Deutsches Elektronen Synchrotron (DESY), Platanenallee 6, Zeuthen, Germany}

\author{Igor~F.~Herbut}
\affiliation{Department of Physics, Simon Fraser University, Burnaby, British Columbia, Canada V5A 1S6}

\author{Michael~M.~Scherer}
\affiliation{Institute for Theoretical Physics, University of Cologne, Z\"ulpicher Stra\ss e 77, Cologne, Germany}

\begin{abstract}
The abelian Higgs model is the textbook example for the superconducting transition and the Anderson-Higgs mechanism, and has become pivotal in the description of deconfined quantum criticality. We study the abelian Higgs model with $n$ complex scalar fields at unprecedented four-loop order in the $4-\epsilon$ expansion and find that the annihilation of the critical and bicritical points occurs at a critical number of
$n_c \approx  182.95\left(1 - 1.752\epsilon + 0.798 \epsilon^2 + 0.362\epsilon^3\right) + \mathcal{O}\left(\epsilon^4\right)\nonumber$.
Consequently, below $n_c$, the transition turns from second to first order. Resummation of the series to extract the result in three-dimensions provides strong evidence for a critical $n_c(d=3)$ which is significantly below the leading-order value, but the estimates for $n_c$ are widely spread.
Conjecturing the topology of the renormalization group flow between two and four dimensions, we obtain a smooth interpolation function for $n_c(d)$ and find $n_c(3)\approx 12.2\pm 3.9$ as our best estimate in three dimensions.
Finally, we discuss Miransky scaling occurring below $n_c$ and comment on implications for weakly first-order behavior of deconfined quantum transitions. We predict an emergent hierarchy of length scales between deconfined quantum transitions corresponding to different $n$.
\end{abstract}

\maketitle

\section{Introduction}

The abelian Higgs (AH) model is one of the most fundamental field theories in both condensed matter and particle physics. It serves as the prime textbook example for the superconducting transition and the Anderson-Higgs mechanism~\cite{zinn1996quantum,peskin2018introduction,herbut2007modern,altland2010condensed}\footnote{To honor its many discoverers appropriately it should  always be referred to as the Anderson-Englert-Brout-Higgs-Guralnik-Hagen-Kibble mechanism~\cite{PhysRev.130.439,PhysRevLett.13.321,PhysRevLett.13.508,PhysRevLett.13.585}, of course.}.
The AH model features a complex scalar field coupled to a fluctuating U(1) gauge field, and it displays two distinct phases separated by a sharp transition: the symmetric phase and the phase with spontaneously broken symmetry.
In the context of superconductors, the symmetric phase is related to the normal metallic state and the spontaneously symmetry broken phase to the superconducting Meissner state.
For the three-dimensional AH model with a single complex scalar, it has been established employing duality arguments that the transition is related to the one of the XY model~\cite{PhysRevLett.47.1556,PhysRevLett.73.1975}, which is known to be continuous, i.e. exhibiting a critical point.
Numerical simulations of lattice versions of both models confirm the conjectured mapping, and also suggest a continuous transition~\cite{PhysRevLett.47.1556,PhysRevB.28.5378,PhysRevB.60.15307}.\medskip

The field-theoretical analysis of the AH model including its generalized version with $n$ complex scalars has turned out to be surprisingly subtle. The subtlety originates in the presence of fluctuating complex scalars and other fluctuating massless modes -- here, coming from the gauge fields.
Such a scenario is generic and also arises in quantum phase transitions of electronic systems.
Indeed, in the AH model, the determination of the nature of transition as a function of $n$, i.e. the question of whether it is discontinuous or continuous, is a long-standing problem in the theory of critical phenomena~\cite{Halperin1974,PhysRevLett.47.1556,PhysRevB.28.5378,PhysRevB.60.15307,Kolnberger1990,radzihovsky1995self,PhysRevLett.73.1975,Herbut1996,PhysRevLett.78.980,PhysRevLett.78.979,Bergerhoff1996,Bartosch2013,Fejos2017}.
In mean-field approximation, for example, the $n=1$ transition is found to be discontinuous. This is further supported by renormalization group (RG) calculations at one-loop order and in $4-\epsilon$ dimensions, which only for a very large value of $n > n_c\sim 183$ show a critical point~\cite{Halperin1974}.
The next-to-leading order loop expansion, on the other hand, shows a dramatic reduction of the critical $n_c$, but due to the large magnitude of this correction, it remains rather inconclusive~\cite{Herbut1996,PhysRevLett.78.980} and has led to the belief that a perturbative RG is not well-suited for the question at hand.
Therefore, despite the tremendous recent advances in higher-loop calculations for (quasi)relativistic field theories~\cite{Kompaniets2017,Batkovich:2016jus,Kompaniets:2016hct,Gracey:2016mio,Zerf:2016fti,Mihaila:2017ble,Zerf:2017zqi,Gracey:2018qba,Gracey:2018ame,Ihrig:2018ojl,Zerf:2018csr,Zerf:2019hhh}, further efforts to improve the description of the $n$-component Abelian Higgs model have not been undertaken for many years.
In this work, we fill this long-standing void by performing a perturbative RG calculation to four-loop order and in the parameter $\epsilon=4-d$, primarily to provide an improved estimate for the critical number of complex scalars $n_c$ above which the transition becomes continuous.

In fact, the $n$-component extension of the AH model is not only a playground for theoretical methods, but has been suggested to be the relevant class of models to describe certain quantum phase transitions beyond the Landau-Ginzburg paradigm~\cite{senthil2004deconfined}. Such transitions have been argued to appear in two-dimensional spin models at zero temperature and in three-dimensional classical loop models between two competing ordered phases and are putatively separated by a critical point -- the deconfined quantum critical point (DQCP).
Explicitly, for $n=2$, the 2+1 dimensional AH model was discussed in this context  due to its relation to the NCCP${}^1$ model~\cite{senthil2004deconfined,senthil2004quantum,senthil2005deconfined,motrunich2004emergent,PhysRevLett.108.137201}. Similar scenarios have been devised for $n=3,4$~\cite{PhysRevB.88.220408,PhysRevB.84.054407} and can possibly be generalized to even higher $n$.
The excitement about the DQCP scenario stems from its unusual phenomenology~\cite{senthil2004deconfined} which -- besides featuring a critical point -- includes the emergence of higher symmetry and fractionalized degrees of freedom becoming deconfined at the critical point.
Indeed, numerical studies of spin models have found indications for such phenomenology, but have also experienced complications due to apparent drift of the values of the critical exponents.
More recent numerical analyses~\cite{PhysRevX.5.041048,PhysRevB.99.195110} argued in favor of a weakly first-order transition instead of a continuous one, but still see emergent symmetry as predicted by the DQCP scenario.
Absence of a continuous transition with the scaling dimensions as found in Monte Carlo simulations is further supported by bounds from symmetry-enhanced conformal field theories~\cite{PhysRevLett.117.131601,RevModPhys.91.015002}.
These combined results suggest that the DQCP scenario is more subtle than initially expected from basic arguments.

The finding that similar scaling properties and violations are present in a variety of models suggests that the numerically observed phenomenology is not uniquely tied to a particular lattice realization and also not exclusively explicable by a fourfold anisotropy of the valence bond solid order parameter~\cite{PhysRevX.5.041048,PhysRevB.99.195110,PhysRevB.80.180414,PhysRevLett.111.087203,PhysRevLett.111.137202,PhysRevLett.101.155702,PhysRevLett.101.167205,PhysRevB.80.045112}.
Rather, it can be expected to be related to a general mechanism which is tied to universal physics described by an effective theory at the phase transition, i.e. the AH model.

In fact, weakly first-order transitions naturally appear in complex conformal field theories~\cite{PhysRevX.5.041048,gorbenko2018walking,PhysRevX.7.031051,PhysRevB.99.195110} where a RG fixed point is complex-valued and, if imaginary parts are small, the complex fixed point slows down and controls the RG flow of real and unitary gapped physical theories.
Such behavior naturally occurs in models where two RG fixed points collide, annihilate and move into the complex plane~\cite{10.1143/PTP.105.809,PhysRevB.71.184519,PhysRevD.80.125005,Gies2006,PhysRevLett.113.106401,PhysRevD.94.025036,PhysRevD.98.096014}, e.g., the AH model below $n_c$.

These observations suggest that the $n_c$ of the three-dimensional $n$-component AH model should be above but still near $n=2,3,4$, to be compatible with the numerical findings.
Through the explicit higher-loop RG analysis presented in this work, we provide the quantitative background for this scenario.
Furthermore, we note that the special point of $n=1$, which exhibits a continuous transition, is not continuously connected to the critical point at large $n>n_c$~\cite{PhysRevX.5.041048}.

The rest of the paper is organized as follows: We summarize our key results in Sec.~\ref{sec:key}. Then, in Sec.~\ref{sec:model}, we introduce the model, notation and also the renormalization group procedure. Sec.~\ref{sec:fps} discusses the renormalization group fixed points and the mechanism of annihilation of two fixed points which underlies weakly first-order transitions. The epsilon expansion at four loops of the critical number of complex scalars is also presented there. We discuss resummation of the series in epsilon in Sec.~\ref{sec:resum} and interpolation between two and four dimensions in Sec.~\ref{sec:interpol}. Finally, we discuss implications of our findings on the DQCP scenario in Sec.~\ref{sec:DQCP} and conclude in Sec.~\ref{sec:conc}. Technicalities are given in the appendix.
%

\section{Key results} \label{sec:key}

We study the $n$-component AH model at four loops and in $4-\epsilon$ dimensions.
An annihilation of the critical and bicritical points occurs at
\begin{align}\label{eq:1}
\hspace{-0.13cm}n_c \approx  n_{c,0}\!\left(1 - 1.752\epsilon + 0.798 \epsilon^2 + 0.362\epsilon^3\right) + \mathcal{O}\left(\epsilon^4\right)\!,
\end{align}
with $n_{c,0}\approx 182.95$.
For $n < n_c$, the phase transition turns from second to weakly first order.
Based on conjectures about the topology of the RG flow between two and four dimensions, we obtain a smooth interpolating function $n_c(d)$.
Our best estimate for the three-dimensional case yields $n_c(3)\approx 12.2\pm 3.9$.
Slightly below $n_c$, Miransky scaling occurs exhibiting a correlation length which is exponentially enhanced upon decreasing the distance to the point of fixed-point collision, e.g., by changing $n$.
Due to the exponential dependence, small variations of $n$ are expected to lead to large changes in the correlation length unless the argument in the exponential is an exceptionally flat function of $n$.
Our calculations do not provide indications of such behavior and we therefore predict a substantial hierarchy of correlation lengths for different $n$ near $n_c$ which may be applicable to $n=2,3,4$ and should be observable in future numerical simulations.

\section{Model and method} \label{sec:model}

The $n$-component AH model is also known as scalar quantum electrodynamics (QED) and is defined in $d$-dimensional euclidean space(time) by the Lagrangian
\begin{align}
  \mathcal{L} = | D_\mu \phi |^2 + \frac{1}{4} F_{\mu\nu}^2+ r |\phi|^2 + \lambda (|\phi|^2)^2\,.
\label{eq: AH model}
\end{align}
Here, $\phi=(\phi_1,...,\phi_n)$ describes the $n$-component complex scalar field with mass term $r$ and quartic interaction $\lambda$. It is minimally coupled to the dynamical non-compact U(1) gauge field $A_\mu$
via the covariant derivative $D_\mu = \partial_\mu - i e A_\mu$ with charge $e$ and indices $\mu,\nu$ run from $0$ to $d-1$. Common summation convention over repeated indices is implied.
The gauge field comes with the field strength tensor $F_{\mu\nu} = \partial_\mu A_\nu - \partial_\nu A_\mu$ and we add a gauge fixing term
$\mathcal{L}_\text{gf} = -\frac{1}{2\xi}(\partial_\mu A_\mu)^2$
where $\xi$ denotes the gauge fixing parameter.

For the case of a single complex scalar field, $n=1$ and in three spatial dimensions, the model is paradigmatically used to describe the superconducting transition~\cite{Ginzburg:1950sr} and also the nematic-to-smectic transition in liquid crystals~\cite{Halperin1974}.
Generally, the scalar mass parameter $r$ can be considered as the tuning parameter of the transition towards, e.g., the superconducting state where the amplitude fluctuations of $\phi$ become massive and -- if the charge is finite $e\neq 0$ -- the phase fluctuations of $\phi$ can be completely absorbed into the gauge field which becomes massive, too. This is the abelian and simplest version of the Higgs mechanism.

\subsection{Renormalization group procedure}\label{sec:rg}

The AH model has an upper critical dimension of $d_c^+=4$ where the charge and the quartic coupling, $g_i \in \{g_1=\alpha = e^2, g_2=\lambda \}$, become marginal simultaneously, i.e. it is perturbatively renormalizable in $d\leq d_c^+$.
Here, we present a perturbative renormalization group analysis in $4-\epsilon$ dimensions using dimensional regularization (DREG) and the modified minimal subtraction scheme ($\overline{\text{MS}}$) to four-loop order.
To that end, we introduce the bare Lagrangian, which is the one from Eq.~\eqref{eq: AH model} with fields and couplings replaced by their bare counterparts, i.e. $x\mapsto x_0$ and $x \in \{\phi, A_\mu, e, r, \lambda, \xi\}$.
Then, the renormalized Lagrangian reads
\begin{align}
	\mathcal{L}' = & Z_\phi |D_\mu \phi|^2+Z_{\phi^2}r\mu^2|\phi|^2+Z_{\phi^4}\lambda\mu^\epsilon (|\phi|^2)^2\\
	\nonumber
	&+ \frac{Z_A}{4}F_{\mu\nu}^2-\frac{1}{2\xi}(\partial_\mu A_\mu)^2\,.
\end{align}
with $D_\mu \phi = (\partial_\mu - i e \mu^{\epsilon/2}A_\mu)\phi$ and $\mu$ defines the energy scale parametrizing the renormalization group flow. Here, we have introduced explicit $\mu$ dependencies to rescale the dimensionless couplings in $4-\epsilon$ dimensions as well as the wavefunction renormalizations $Z_\phi$ and $Z_A$ to relate the bare and the renormalized Lagrangian through $\phi_0=\sqrt{Z_\phi}\phi$ and $A_{0,\mu}=\sqrt{Z_A}A_\mu$.
Accordingly, the bare and the renormalized bosonic mass terms are related by $r=r_0\mu^{-2}Z_\phi Z_{\phi^2}^{-1}$ and we obtain the following relations between the bare and the renormalized couplings
\begin{align}
	\alpha = e_0^2\mu^{-\epsilon}Z_A\,,\quad \lambda = \lambda_0 \mu^{-\epsilon}Z_{\phi}^2 Z_{\phi^4}^{-1}\,.
\end{align}
For completeness, we note that the flow of the gauge-fixing parameter is encoded in the relation $\xi=\xi_0 Z_A^{-1}$.
The calculation of the renormalization constants $Z_x$ with $x \in \{\phi, \phi^2, \phi^4, A\}$ at four-loop order is  performed by using an automated protocol which is described in App.~\ref{sec:toolchain}.

From the renormalization contants, we can construct the renormalization group beta functions which are defined as the logarithmic derivatives of the dimensionless renormalized couplings $\{g_1=\alpha = e^2, g_2=\lambda \}$ with respect to $b=\mu^{-1}$. Schematically, the beta functions have the form
\begin{align}
  \beta_{i} = \frac{\mathrm{d}g_i}{\mathrm{d}\ln b} = \epsilon g_i + \sum_k\beta_{i}^{(k\ell)}\,,
\end{align}
where the index $k$ indicates the loop order. Explicitly, we obtain for the gauge coupling $\alpha$,
\begin{align}
  \beta_\alpha^{(1\ell)} &= -\frac{n}{3}\alpha^2\,,\\[5pt]
  \beta_\alpha^{(2\ell)} &= -2 n \alpha^3\,,\\[5pt]
  \beta_\alpha^{(3\ell)} &= \textstyle \left(\frac{49}{72} n^2-\frac{29}{8} n\right)\alpha ^4 -\frac{n^2+n}{2}\alpha ^3 \lambda +\frac{n^2+n}{8}\alpha ^2 \lambda ^2\,.
\end{align}
The beta functions for the quartic self-interaction coupling $\lambda$ up to three-loop order are given accordingly as
\begin{align}
\beta_\lambda^{(1\ell)} &= -6 \alpha ^2+6 \alpha  \lambda -(n+4)\lambda ^2\,,\\[5pt]
   \beta_\lambda^{(2\ell)} &=\textstyle \left(\frac{14}{3} n+30\right)\alpha ^3 - \left(\frac{71}{6} n+\frac{29}{2}\right)\alpha ^2 \lambda \\
   &\textstyle - (4 n+10)\alpha  \lambda ^2+\left(\frac{9}{2} n+\frac{21}{2}\right)\lambda ^3\,, \nonumber\\[5pt]
   \beta_\lambda^{(3\ell)} &=\textstyle \left(-\frac{7 }{18}n^2+\left[\frac{203}{8}-27 \zeta _3\right] n+\frac{367}{8}-45 \zeta _3\right)\alpha ^4\\
      & \textstyle + \left(-\frac{5 }{216}n^2+\left[18 \zeta _3-\frac{989}{8}\right] n-\frac{889}{4}-54 \zeta _3\right)\alpha ^3 \lambda \nonumber \\
      & \textstyle +\left(\frac{43 }{16}n^2+\left[18 \zeta _3+\frac{1749}{16}\right] n+\frac{1093}{8}+126 \zeta _3\right)\alpha ^2 \lambda ^2 \nonumber \\
      & \textstyle +\left(-\frac{33 }{16}n^2+\left[-15 \zeta _3-\frac{461}{16}\right] n-\frac{185}{4}-33 \zeta _3\right)\lambda ^4 \nonumber \\
      & \textstyle +\left(\left[\frac{25}{2}-6 \zeta _3\right] n+\frac{29}{2}+6 \zeta _3\right)\alpha  \lambda ^3\,, \nonumber
\end{align}
where $\zeta_3 = \zeta(3)$ denotes the Riemann $\zeta$-function.
The four loop contributions are presented in App.~\ref{sec:beyondoneloop}.
Note that the beta functions are gauge parameter independent as expected.
Further, the expressions are in full agreement to the two-loop results \cite{Kolnberger1990,herbut2007modern}.
Also, in the limit $\alpha \rightarrow 0$ we recover the beta functions of the purely bosonic $2n$-vector model up to four-loop order~\cite{Kleinert}.
For completness, we list the beta function of the gauge fixing parameter in the App.~\ref{sec:beyondoneloop}. Note that we also provide the field anomalous dimensions there.

\section{Fixed points and their collision}\label{sec:fps}

Generally, fixed points (FP) of the RG flow are determined by the solutions of the beta functions
\begin{align}
  \beta_i(\{ g_i^\ast \}) = 0\ \ \forall\ i\,,\label{eq: fixed point condition}
\end{align}
where $\{g_i\}$ denotes the set of couplings and $\{g_i^\ast\}$ the set of fixed point coordinates.
Linearization of the RG flow near a FP determines its critical behavior,
\begin{align}
  \beta_i(\{g_i\}) = \sum\limits_{j}B_{ij}(g_j - g_j^*) + \mathcal{O}(\{g_i^2\})\,,
\end{align}
where $B_{ij} = \partial \beta_i/ \partial g_j$  defines the stability matrix. The eigenvalues $\theta_i$ and eigenvectors $\vec{v}_i$ of $B_{ij}$ determine the critical exponents and the corresponding directions. Positive eigenvalues describe relevant and negative ones irrelevant directions. A vanishing eigenvalue means the RG flow in this direction becomes marginal.

When two FP solutions coincide, the flow between them is marginal, implying that the stability matrix has a vanishing eigenvalue, i.e. the determinant vanishes,
\begin{align}\label{eq:det}
  \det (B_{ij})|_{\{g_i^*\}} = 0\,,
\end{align}
providing an additional condition to the fixed point equations in Eq.~\eqref{eq: fixed point condition}. We use Eq.~\eqref{eq:det} and the condition of a vanishing eigenvalue, $\theta_i=0$ for some $i$, as a criterion for the appearance of a fixed point collision for both, the abelian Higgs model in $4-\epsilon$ dimensions as well as the non-linear sigma model in $2+\epsilon$ dimensions, see below.

\subsection{AH model and fixed point collision}\label{sec: one loop}

\begin{figure}[b!]
\includegraphics[width=\columnwidth]{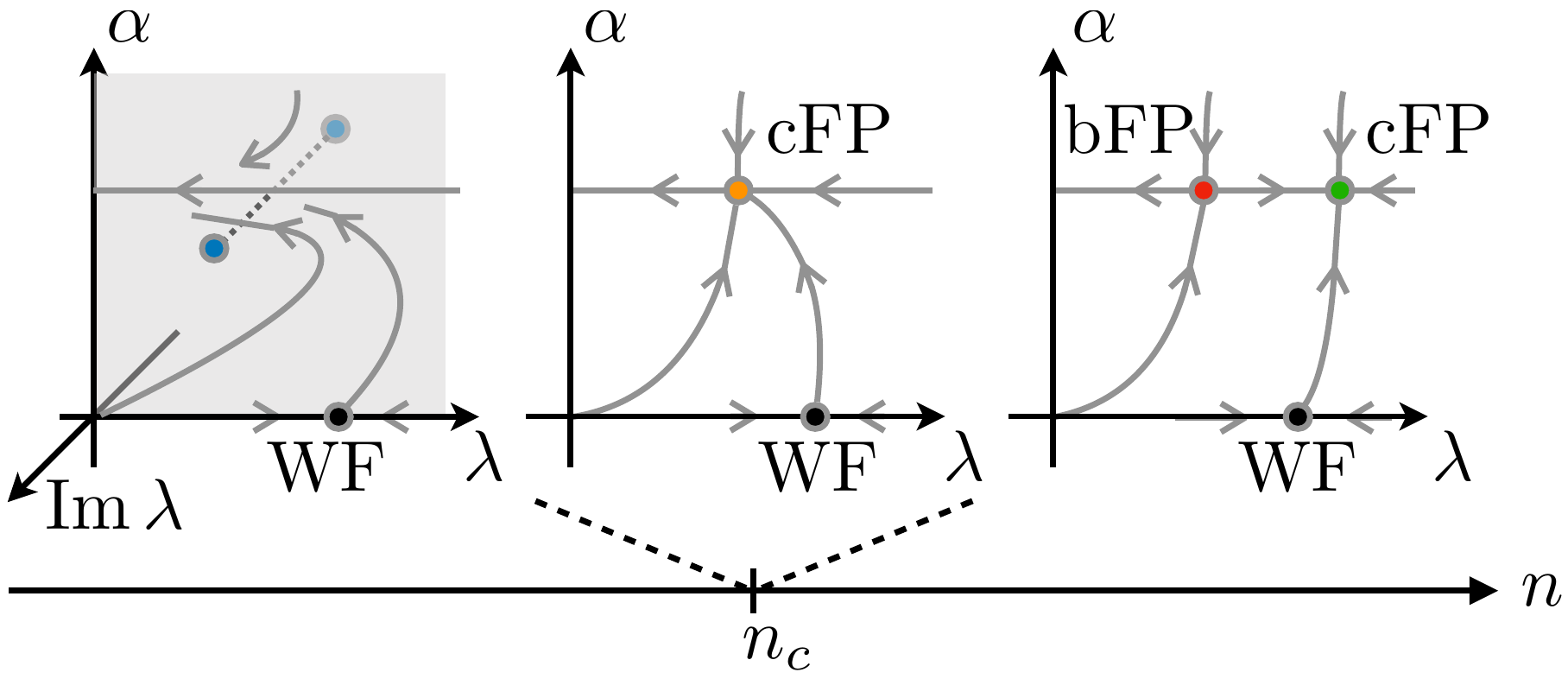}
  \caption{Renormalization group flow and fixed point annihilation in the Abelian Higgs model. The perturbative computations suggest that
  above a critical number of scalar compenents $n_c$ an IR-attractive charged FP exists. At $n_c$ this
  collides with the bicritical FP so that they annihilate and disappear in the complex plane of the quartic coupling.}
  \label{fig: fixed point collision}
\end{figure}

At one-loop order, the beta functions read
\begin{align}\label{eq:beta1loop}
  \beta_\alpha^{(1\ell)} = -\frac{n}{3}\alpha^2\,,\  \beta_\lambda^{(1\ell)} &= -6 \alpha ^2+6 \alpha  \lambda -(n+4)\lambda ^2\,.
\end{align}
This set of beta functions features four fixed points (FP) where $\epsilon$, i.e. $\beta_i(\alpha^*,\lambda^*)=0$ and scaling may emerge.
In $d=4-\epsilon$, one FP is the trivial Gaussian fixed point, $\alpha^\ast=\lambda^\ast=0$, and another one is the Wilson-Fisher fixed point of the scalar O($2n$) model, $\alpha^\ast=0, \lambda^\ast=\epsilon/(n+4)+ \mathcal{O}(\epsilon^2)$.

The two remaining non-trivial non-Gaussian fixed points have identical FP coordinate of the gauge coupling, $\alpha^* = 3\epsilon/n + \mathcal{O}(\epsilon^2)$.
There are two corresponding FP solutions for the quartic interaction,
\begin{align}
  \lambda^*_\pm = \frac{3 (18+n \pm \sqrt{s})}{2n(n +4)}  \epsilon + \mathcal{O}(\epsilon^2)\,, \label{eq:quarticFP}
\end{align}
where $s=n^2-180 n-540$. We refer to ($\alpha^\ast$,$\lambda^*_+$) as the charged FP  and to ($\alpha^\ast$,$\lambda^*_-$) as the bicritical FP . The charged FP is irrelevant in both directions, i.e. represents a stable FP. The bicritical FP is unstable to perturbations in the quartic coupling.

The fixed points have a strong dependence on the
number of field components $n$, in particular, the quartic coupling only provides real-valued solutions above a certain critical $n > n_c$, when the radicand $s$ of the square root in Eq.~\eqref{eq:quarticFP} is positive.
At $n =n_c$ the two fixed points collide and, for $n < n_c$, drift  into the complex plane.
The one-loop analysis predicts that this happens for values of
\begin{align}
  n_{c,0} = 6 (15 + 4 \sqrt{15}) \approx 182.95\,. \label{eq: nc0}
\end{align}
Consequently, for $n \geq n_c$ there is a continuous transition while below  it is expected to become first-order~\cite{Halperin1974}. Note that we neglected an unphysical negative solution $n_{c,-}= 6 (15 - 4 \sqrt{15}) +\mathcal{O}(\epsilon) \approx -2.952+\mathcal{O}(\epsilon)$. At the next order around the negative solution $n_{c,-}$, we find it to become even more negative
$n_{c,-} \approx -2.952(1 + 1.62323 \epsilon +\mathcal{O}(\epsilon^2))$.

\subsection{Fixed-point collision at four loops}

In the previous section, we considered the fixed points of the AH model at one-loop level and found that two fixed points collide and disappear in the complex plane when tuning the number of scalar components $n$. The critical number $n_c$ below which the fixed points
disappear in the complex plane was computed by analyzing the imaginary part of the quartic coupling's charged FP coordinate. Here, we compute corrections to this value at higher orders.

\begin{figure}[b!]
\begin{minipage}[t]{0.65\linewidth}
  \vspace{0pt}
  \includegraphics{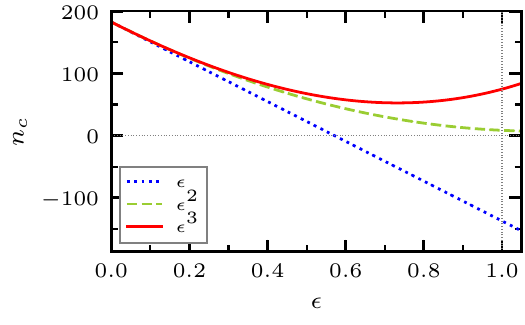}
 \end{minipage}
 \hfill
 \begin{minipage}[t]{0.3\linewidth}
   \vspace{2pt}
 \begin{tabular*}{0.9\linewidth}{@{\extracolsep\fill}lr}
    \hline\hline
     Order    & $\epsilon=1$ \\
    \hline
    $\epsilon^0$ & 182.95  \\
    $\epsilon^1$ & -137.54  \\
    $\epsilon^2$ & 8.42   \\
    $\epsilon^3$ & 74.69 \\
     \hline\hline
  \end{tabular*}
  \end{minipage}
   \caption{Direct evaluation of the epsilon expansion in Eq.~\eqref{eq:ncseries} for the critical number of components $n_c$ at different orders in $\epsilon$.}
  \label{noresumplots}
\end{figure}

We use Eqs.~\eqref{eq: fixed point condition} and \eqref{eq:det} to calculate the fixed point collision point $n_c$ in the parameter $n$ at higher orders in the $\epsilon$ expansion to obtain the series expansion
\begin{align}\label{eq:ncseries}
  n_c \approx n_{c,0}+ n_{c,1}\epsilon + n_{c,2}\epsilon^2 + n_{c,3}\epsilon^3+ \mathcal{O}(\epsilon^4)\,,
\end{align}
with $n_{c,0}$ from Eq.~\eqref{eq: nc0} and we find
\begin{align}
	n_{c,1} =&-\frac{9}{70}\left(317 \sqrt{15}+1265\right)\,,\\[10pt]
	n_{c,2} =&\frac{53396968+25893277 \sqrt{\frac{3}{5}}}{768320}\nonumber\\
	+&\frac{27981-22872 \sqrt{\frac{3}{5}}}{245} \zeta_3\,,\\[10pt]
	n_{c,3} =&\frac{197218191162096-49532015359609 \sqrt{15}}{189744307200}\nonumber\\
	+&\frac{\left(9327- 7624\sqrt{\frac{3}{5}}\right)}{9800}\pi ^4+\frac{1251897}{490}\zeta_5-\frac{492241}{49 \sqrt{15}}\zeta_5\nonumber\\
	+&\frac{\left(5103190199 \sqrt{15}-17908675920\right)}{50421000} \zeta_3\,,
\end{align}
where, again, $\zeta_3 = \zeta(3)$ and $\zeta_5 = \zeta(5)$ denote the Riemann $\zeta$~function. Numerical evaluation then provides the approximate expression presented in Eq.~\eqref{eq:1}.
This series expansion exceeds previous estimates by two orders in $\epsilon$. Since the coefficients in the series of Eq.~\eqref{eq:ncseries} are still decreasing in magnitude, we provide direct evaluation for $\epsilon=1$, see Fig.~\ref{noresumplots}. In the following, we will explore resummation of the series.

\normalsize

\section{Resummation of $4-\epsilon$ expansion}\label{sec:resum}

\begin{figure}[b!]
  \includegraphics{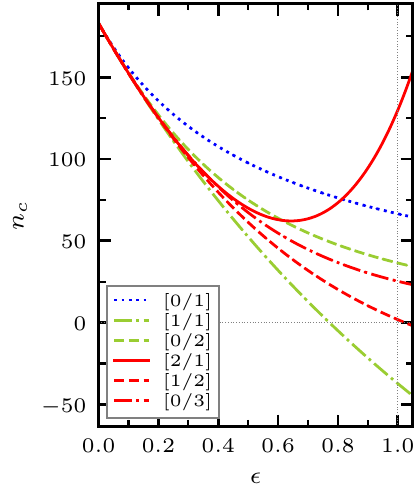}\hfill
 \includegraphics{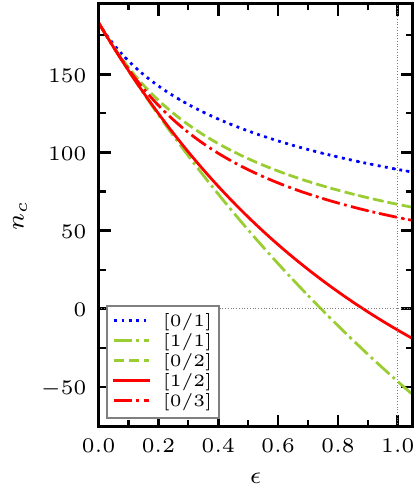}
  \caption{Resummed epsilon expansion for the critical number of components $n_c$. Left panel: Pad\'e approximants. Right panel: Pad\'e-Borel approximants.}
  \label{resumplots}
\end{figure}

\begin{table}[b!]
  \caption{Pad\'e and Pad\'e-Borel approximants of the  $\epsilon$ expansion evaluated at $d=3$ ($\epsilon=1$). The Pad\'e-Borel $[2/1]$ is
  not availabe due to a singularity in the integral of the Borel sum.}
  \label{tab: Pade nc}
  \begin{tabular*}{\linewidth}{@{\extracolsep\fill}lrrrr}
    \hline\hline
    Order   &  & $[m/n]$ &     Pad\'e       &     Pad\'e-Borel      \\
    \hline
    $\epsilon$		 &           &$[0/1]$        &  66.48        &  89.11      \\
    $\epsilon^2$		        &         & $[0/2]$        &  36.42        &   66.78 \\
                       & & $[1/1]$        &  --37.25       &   --46.53 \\
    $\epsilon^3$	 &   &$[0/3]$        &  25.27        &   58.42  \\
                       & &$[2/1]$        &  129.81       &   -- \\
                       &  &$[1/2]$        &  1.81         &   --13.65 \\
    \hline\hline
  \end{tabular*}
\end{table}

The expansions from perturbative RG calculations are asymptotic series. Consequently, obtaining reliable estimates for $n_c$ at $d=3$ requires resummation.
Whether or not such resummation yields reliable results depends on the underlying model, the order of the expansion and the knowledge about the large-order behavior.
Here, we explore three resummation schemes for the series in $4-\epsilon$ dimensions, i.e. 1.~Pad\'e approximants, 2.~Pad\'e-Borel approximants and 3.~Borel resummation.
The employed schemes presented here are conventional and we use standard notation, e.g., for the Pad\'e approximants $[m/n]$.
Explicit definitions are provided in App.~\ref{app:pade}. We display the available Pad\'e and Pad\'e-Borel approximants order by order in Fig.~\ref{resumplots} and in Tab.~\ref{tab: Pade nc}.
The Borel resummation scheme including the corresponding set of resummation parameters is introduced in App.~\ref{eq:borel} and we merely state the results here.
Optimization through variation of the resummation parameters according to the principle of minimal sensitivity and the principle of fastest convergence yields a negative weighted mean value for $n_c(\epsilon=1)$ with a huge error of $n_c(\epsilon=1)\approx -52 \pm 45$.
In the present case, we find that the three methods do not allow us to extract a precise numerical result for $n_c(\epsilon=1)$ and estimates are scattered over a rather large range.

The main conclusion drawn from this one-sided resummation analysis is therefore merely that $n_c(\epsilon\!=\!1)$ lies significantly below the leading-order result of $n_{c,0}\approx 183$.

\section{Dimensional interpolation}\label{sec:interpol}

In view of the high order series for $n_c$ in $\epsilon$, we would like to obtain a better estimate on $n_c(\epsilon=1)$ than the one from the resummations explored in the previous section. This is possible through an interpolation between two and four dimensions and by additionally including the assumption of
$n_c(d=2+\epsilon) = a \epsilon$, with unknown coefficient $a >0$.
Before we carry out the explicit interpolations between two and four dimensions, we briefly review the arguments leading to that  conjecture.

\subsection{Nonlinear $\sigma$ model in $2+\epsilon$ dimensions}

The first item is suggested from studying the nonlinear sigma model in $2+\epsilon$ dimensions, e.g., in the $\mathbb{CP}^{n-1}$ formulation~\cite{Hikami1979,March-Russell1992}, which can be related to the abelian Higgs model in the following way:
Upon additional expansion in large $n$ the $2+\epsilon$ expansion of the nonlinear sigma model and the $4-\epsilon$ expansion of the AH model describe the same fixed point~\cite{March-Russell1992}. This is exhibited by comparing their critical exponents with the large-$n$ result expanded in the respective dimension which turn out to be identical. In App.~\ref{app:cp1}, we provide further evidence for this correspondence by exploiting the four-loop expansion in $4-\epsilon$ dimensions for the correlation length exponent for the first time.
Therefore, the two expansions near two and four dimensions in the respective models can be employed to provide for a continuous interpolation of the same renormalization group fixed points in the plane $(n,d)$ for large enough $n > n_c(d)$.

In the $\mathbb{CP}^{n-1}$ model $n_c(d\rightarrow 2)\rightarrow 0$~\cite{Hikami1979,March-Russell1992}.
Moreover in $d=2+\epsilon$, the beta function of the model coupling $t$ features a real valued fixed-point solution $t^\ast$ for all $n>0$, which for small $\epsilon$ is $t^\ast=\epsilon/n+\mathcal{O}(\epsilon^2)$.
Therefore, a fixed-point collision is expected to be exhibited in the symmetry-allowed higher-derivative terms of the non-linear sigma model.
For example, the RG scaling of the canonically least irrelevant higher-derivative terms, i.e. the four-derivative terms, is found to be\footnote{The scaling of higher-derivative terms is found to be $y=2-2k+\epsilon+2k(k-1)\, t^\ast$ for  terms $2k$ derivatives, cf. Ref.~\cite{Lerner1990}.} $y_{4}=-2+\epsilon+4\, t^\ast$ at one-loop order. A fixed-point collision would now be indicated by a vanishing of $y_4$, i.e. when the scaling becomes marginal, cf.~Eq.~\eqref{eq:det}, which at that order yields $n_c(d=2+\epsilon) = 2\epsilon+\mathcal{O}(\epsilon^2)$.

At this point a word of caution seems in order. Concerning a possible fluctuation-induced fixed-point destabilization due to the RG relevance of terms with an even larger number of derivatives in nonlinear sigma and related models, there has been a extended discussion in the literature, e.g., Refs.~\cite{Lerner1990,PhysRevLett.71.384,PhysRevB.55.R10169}, which we here will not attempt to resolve. Instead, we take on a pragmatic approach by assuming that near two dimensions the first fluctuation-induced relevant direction comes from the least irrelevant canonical terms, which is the four-derivative term. Moreover, since $t^\ast$ is of $\mathcal{O}(1)$ at that point and therefore not small, the corrections to the leading-order behavior of $y_4$ cannot be expected to be small either. We therefore refrain from fixing the coefficient of the epsilon expansion of $n_c(d+\epsilon)$ and just write $n_c(d=2+\epsilon) = a \epsilon$, with unknown coefficient $a >0$ as stated above, see also Ref.~\cite{PhysRevX.5.041048}.

\subsection{Interpolation}\label{sec:inter}

A suitable interpolation between the two critical dimensions can now be constructed by using a polynomial ansatz.
To that end, we use both epsilon expansions for $n_c$, simultaneously, and set up an interpolating function in the interval $d\in [2,4]$.
More specifically, we choose a polynomial interpolation with polynomial $P_{i, j} (d)$ of degree $i+j$, where $i$ ($j$) denotes the highest order of the epsilon expansion in $d=2+\epsilon$ ($d=4-\epsilon$) dimensions, i.e $i=1$ and $j\in \{1,2,3,4\}$.
We fix the polynomial coefficients of the first $i+1$ terms with the expansion near the lower critical dimension.
The remaining $j+1$ higher-order coefficients are then determined from the requirement that the $j$ lowest derivatives of $P_{i, j} (d)$ at $d=4$ correspond to the $4-\epsilon$-expansion.
The resulting polynomials are then by construction $i$-loop exact near the lower critical dimension and ($j$-loop) exact near the upper critical dimension. From this calculation, we obtain interpolating functions for general coefficient $a$, i.e.
$P_{1,1}(d=3) \approx 11.35 + 0.25 a$, 	$P_{1,2}(d=3) \approx 2.023 + 0.125 a$, $P_{1,3}(d=3) \approx 17.86 + 0.0625 a$,
exhibiting that the dependence on the coefficient $a$ is rather weak.

\begin{figure}[t!]
  \includegraphics{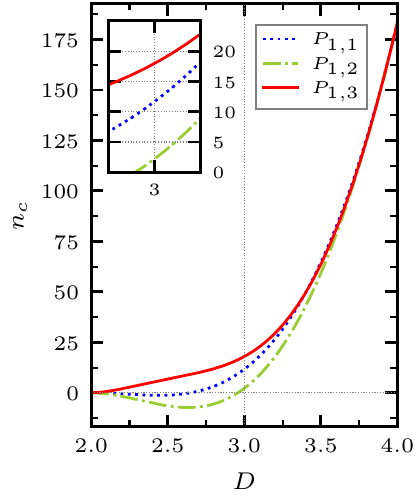}
  \includegraphics{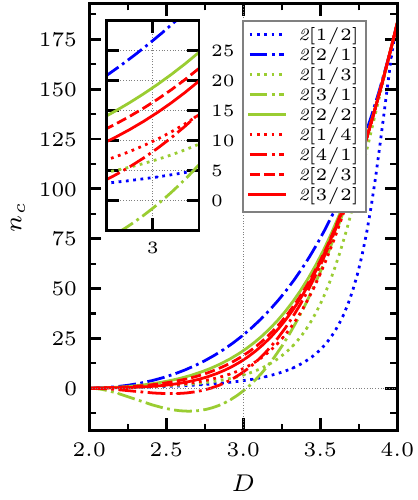}
  \caption{Left panel: Estimates from polynomial interpolation for the critical number of components $n_c(d)$ for $a=1$. Right panel: same for two-sided Pad\'e approximants. }
  \label{fig: interpolation}
\end{figure}

In Ref.~\cite{PhysRevX.5.041048} the additional conjecture that $n_c(d)$ increases monotonically, i.e. $n_c'(d)>0$, has been put forward, however, without further justification. Here, we briefly discuss the impact of this assumption:
while $P_{1,3}(d)$ increases monotonically for $0<a\lesssim 80$, $P_{1,1}(d)$ requires $10\lesssim a\lesssim 200$ and $P_{1,2}(d)$ is never monotonous for $2<d<4$. We show the results for the polynomial interpolation as a function of $d$ in the left panel of Fig.~\ref{fig: interpolation} and compiled their estimates at $d=3$ in Tab.~\ref{tab: interpolations}.

Following the same reasoning, we can set up two-sided Pad\'e approximants $\textit{2}_{[m/n]}(d)$, cf. the right panel of Fig.~\ref{fig: interpolation} and App.~\ref{app:pade} for definitions.
Tentatively, setting $a=1$, we observe that increasing the order in $\textit{2}_{[m/n]}(3)$ from $m+n=3$ to $m+n=5$ decreases the range spanned by the estimates for $n_c(d=3)$ order by order. This behavior is even more pronounced upon disregarding the values which belong to non-monotonous interpolations, i.e. dropping the values for $\textit{2}[3/1]$ and $\textit{2}[4/1]$.

We use the highest order, $m+n=5$, including the non-monotonous $\textit{2}[4/1]$, to calculate the average and error
\begin{align}\label{eq:best}
	n_c(d=3)\approx 12.2\pm 3.9\,.
\end{align}
This is our best estimate based on the above reasoning. We have also studied the dependence on the parameter $a$ and find that, within an extended range of $a \in [0.2,5]$, this estimate varies mildly, see Tab.~\ref{tab: a dep twosided Pade} in App.~\ref{app:pade}.

\begin{table}[t!]
  \caption{Estimates from polynomial interpolation and two-sided Pad\'e approximation for the critical number of components $n_c$ at $d=3$ dimensions for $a=1$. The values which belong to a non-monotonous  interpolating function $n_c(d)$ for $d\in [2,4]$ are printed in italics.}
  \label{tab: interpolations}
  \begin{tabular*}{\linewidth}{l@{\extracolsep\fill}rlr}
    \hline\hline
    polynom. Int. & & Two-sided Pad\'e & \\
    \hline
    $P_{1,1}$ & \textit{11.60} & $\textit{2}[2/1]$ & 26.62 \\
     &  & $\textit{2}[1/2]$ & 3.77 \\
     \hline
     $P_{1,2}$ & \textit{2.15} & $\textit{2}[2/2]$ & 18.76 \\
     & & $\textit{2}[1/3]$ & 6.56 \\
     & & $\textit{2}[3/1]$ & \textit{-1.40} \\
     \hline
     $P_{1,3}$ & 17.92 & $\textit{2}[2/3]$ & 16.32 \\
     & & $\textit{2}[3/2]$ & 14.11 \\
     & & $\textit{2}[1/4]$ & 9.80 \\
     & & $\textit{2}[4/1]$ & \textit{8.06} \\
    \hline\hline
  \end{tabular*}
\end{table}

\section{Deconfined pseudo criticality}\label{sec:DQCP}

The $n$-component abelian Higgs model has been argued to describe the universal properties of the N\'eel--valence-bond-solid transition~\cite{senthil2004deconfined,senthil2004quantum} and related scenarios.
Recent numerical simulations of this transition revealed strong violations of scaling and have argued that this transition appears to be weakly first order with an anomalously large correlation length and drifting critical exponents, see Refs.~\cite{PhysRevX.5.041048,PhysRevX.7.031051,PhysRevB.99.195110} and references therein.

\subsection{Miransky scaling}

The behavior described above may be due to emergent walking behavior that appears when two fixed points have just been annihilated and vanished into the complex plane, e.g., just as it happens in the AH model below $n_c$. Here, we briefly recap the underlying reasoning using the beta functions at one-loop order, cf. Eqs.~\eqref{eq:beta1loop}.
At that order for the gauge coupling the fixed point value $\alpha_\ast=3\epsilon/n$ can be acquired for any choice of $n \lessgtr n_c$. Further, the gauge coupling is irrelevant and we therefore can replace $\alpha=\alpha_\ast$ in the beta function for the quartic coupling $\lambda$, to obtain
\begin{align}
	\frac{d\lambda}{d\ln b}\Big|_{\alpha_\ast}=\left(1+\frac{18}{n}\right)\epsilon \lambda-(n+4)\lambda^2 -\frac{54}{n^2}\epsilon^2\,.
\end{align}
This flow equation can be integrated yielding
\begin{align}
\ln\!\left(\!\frac{b_{\mathrm{IR}}}{b_{\mathrm{UV}}}\!\right)\!=\!\frac{-2n}{\epsilon\sqrt{-s}}\arctan\!\left(\frac{2n(4+n)\lambda\!-\!(18+n)\epsilon}{\epsilon\sqrt{-s}}\right)\!\bigg|_{\lambda_{\mathrm{UV}}}^{\lambda_{\mathrm{IR}}},\nonumber
\end{align}
with $s=n^2-180 n-540$, again. For $n\lesssim n_c$, where $s<0$ and $|s|\ll 1$, the flow of $\lambda$ proceeds as follows: $\lambda$ starts at some positive value, goes through a walking regime and eventually diverges towards negative values. Therefore, in the infrared and the ultraviolet, the argument of the $\arctan$ is always large, i.e. we can use $\lim_{x\to \pm \infty} \arctan x=\pm\pi/2$. Arbitrarily choosing $b_{\mathrm{UV}}=1$ and renaming
$b_{\mathrm{IR}}=L_{\mathrm{IR}}$, we obtain the exponentially large infrared length scale~\cite{PhysRevX.5.041048,PhysRevD.80.125005,Gies2006}
\begin{align}\label{eq:mir}
	L_{\mathrm{IR}}(n)\simeq\exp\left(\pi f(n)\right)\,,\ \mathrm{where}\ f(n)=\frac{2n}{\epsilon\sqrt{-s}}\,,
\end{align}
which is also referred to as Miransky scaling. Note that including the running of the infrared-attractive gauge coupling yields power-law corrections to this behavior~\cite{PhysRevD.84.034045}. Further, higher-loop contributions will generally modify the function $f(n)$.

Now, approaching $n_c$ from below, this correlation length grows exponentially to diverge at $n_c$. Therefore, to obtain a large correlation length, one needs to be close to $n_c$, which at one-loop order, Eq.~\eqref{eq:mir}, is $n_{c,0}\sim 183$.
Generically, small changes in $n\lesssim n_c$ induce a large hierarchy in length scales, for example at $\epsilon=1$, $L_{\mathrm{IR}}(180)/L_{\mathrm{IR}}(179)\sim 833$ or $L_{\mathrm{IR}}(179)/L_{\mathrm{IR}}(178)\sim 97$ which, however, decreases rapidly when $n$ moves away from $n_c$. In case, $n$ is further away from $n_c$, the correlation length is, of course, exponentially suppressed.

Including higher-loop contributions into this reasoning will significantly change the quantitative aspects of this behavior, but not its qualitative features. For this work, we refrain from providing a corresponding quantitative analysis, since this would require an appropriate resummation of the underlying higher-loop beta functions. We postpone such an analysis to the future and describe the underlying qualitative implications, instead.

\subsection{Hierarchy of length scales}

In numerical simulations of deconfined quantum transitions, similar scaling violations have been observed for various lattice models~\cite{PhysRevB.80.180414,PhysRevLett.111.087203,PhysRevLett.111.137202,PhysRevLett.101.155702,PhysRevLett.101.167205,PhysRevB.80.045112} and, notably, also for a range of values for $n$, i.e. $n=2,3,4$~\cite{PhysRevB.88.220408,PhysRevB.84.054407}.
Here, based on our four-loop calculations and the interpolating functions from Sec.~\ref{sec:inter}, we will explore whether such scaling violations for $n=2,3,4$ can be consistently be ascribed to the pseudo-critical or walking behavior related to Miransky scaling, cf.~Eq.~\eqref{eq:mir}, in the three-dimensional AH model. First, we note that our estimate of $n_c(3)\approx 12.2\pm 3.9$, Eq.~\eqref{eq:best}, tentatively supports the scenario that $n_c(3)$ is of comparable order as the physically relevant $n=2,3,4$. Uncertainties in this determination are still large and it is therefore conceivable that the true $n_c(3)$ is closer to $n=2,3,4$. The following reasoning, however, will not rely on the precise number $n_c(3)$, but on the behavior of the interpolating function $n_c(d)$ for $d\in [2,4]$ as exhibited in Fig.~\ref{fig: interpolation}.

The essential feature of the infrared length scale as given by the Miransky scaling, Eq.~\eqref{eq:mir}, is that the length scale diverges when approaching $n_c$ from below. This is related to the singularity in the argument $f(n)$ appearing in the exponential at $n_c$.
Sufficiently close to $n_c$, the slope of $L_{\mathrm{IR}}(n)$ diverges, too, and small variations of $n$ are expected to lead to large changes in $L_{\mathrm{IR}}(n)$. This behavior is generic unless the function $f(n)$ is exceptionally flat just before it develops the singularity, e.g., like an inverse step function $\Theta^{-1}(n_c-n)$. Whereas, we cannot quantitatively determine the function $f(n)$ from our present analysis, our interpolating functions for $n_c(d)$, cf. Fig.~\ref{fig: interpolation}, do not provide indications for such exceptionally flat behavior.
We therefore expect that near $n_c$ a change of $n\to n\pm 1$ significantly impacts the correlation length due to its exponential behavior.
On the other hand, if $n$ is further away from $n_c$, it is exponentially suppressed and the pseudo-critical or walking regime where, approximate scaling can be observed, should be small.

We conclude that, generically, a substantial hierarchy of characteristic correlation lengths can be expected for different values of $n$. In particular, if the true $n_c(3)$ lies close enough to $n=2,3,4$, we predict that the three cases should exhibit this large hierarchy of length scales, which can be probed in numerical simulations by comparing the correlation lengths for $n=2,3,4$.

\section{Conclusion and outlook}\label{sec:conc}

We have delivered a study of the $n$-component AH model at four loops and in $4-\epsilon$ dimensions and analysed the critical number of complex scalars $n_c$, below which a fixed-point collision appears and the phase transition turns from second to first order.
Based on a series of assumptions on the topology of the RG flow between two and four dimensions, we have obtained a smooth interpolating function $n_c(d)$ and our best estimate for the three-dimensional case yields $n_c(3)\approx 12.2\pm 3.9$.
Slightly below $n_c$ weakly first order behavior occurs with a correlation length which is governed by Miransky scaling, i.e. it is exponentially suppressed upon increasing the distance to the point of fixed-point collision.

Due to this exponential dependence, even small variations of $n$ are expected to induce substantial changes in the correlation length unless the argument appearing in the exponential is an exceptionally flat function of $n$.
We do not find indications of such behavior and therefore predict a large hierarchy of correlation lengths for different $n$, if $n_c(3)$ is near these values.
This is relevant to the results from different lattice studies of the corresponding deconfined phase transitions for $n=2,3,4$ and we expect that indications of this hierarchy of length scales can be measured in numerical simulations.

In the future, it will be interesting to extract more quantitative estimates for the merging line $d_c(n)$ employing appropriate resummations of the beta functions \cite{Folk1998}.

\paragraph*{Acknowledgments.}
We thank A.~Nahum for correspondence and A.~L\"auchli for discussion.
M.M.S. is supported by the Deutsche Forschungsgemeinschaft (DFG), Projektnummer 277146847 -- SFB 1238 (project C04). This project has received funding from the European Union's Horizon
2020 research and innovation programme under the Marie Sk\l odowska-Curie grant agreement No.~764850 (SAGEX).
This work was supported by the NSERC of Canada.

\appendix

\section{Tool chain} \label{sec:toolchain}

We calculate the wave-function and vertex renormalization constants using dimensional regularization (DREG) within the modified minimal subtraction scheme ($\overline{\text{MS}}$).
In order to obtain the $Z$ factors of specific Green-functions we apply the Infrared Rearrangement procedure (IRR) introduced in Ref.~\cite{Misiak:1994zw,Chetyrkin:1997fm}.
This method allows to extract the overall UV divergence appearing in a Feynman diagram in terms of single scale massive tadpole integrals.
The method is based on an exact decomposition of every massless propagator into a sum of massive propagators (of increased power) dropping irrelevant UV finite terms, only.
Further it avoids contamination from any IR divergences, which in DREG in general can get regulated in terms of $1/\epsilon$ poles, too.
However, the ability to calculate just the overall divergence requires a full subtraction of all existing sub-divergences via an explicit insertion of wave-function and vertex counter terms.

The number of Feynman diagrams for each Green function we have calculated is shown in Tab.~\ref{TAB:DiagramListing}.
To factorize the flavor amplitude on diagram level,
we introduced an auxiliary (non-propagating) field $\sigma$ which helps to split the original flavour-non-factorizeable four scalar vertex into a sum of products of two scalar-scalar-$\sigma$ vertices.

To be able to calculate the large number of diagrams we first use
 \code{QGRAF}~\cite{Nogueira:1991ex} to generate complete sets of them.
Then \code{q2e} and \code{exp}~\cite{Seidensticker:1999bb} are used to map all Feynman diagrams on one-scale massive tadpole integral topologies and to generate diagram source files.
Next, \code{FORM}~\cite{Vermaseren:2000nd,Kuipers:2012rf,Ruijl:2017dtg} is used to process the diagram source files.
It rewrites the amplitudes in terms of massive tadpole integrals with different powers of propagators.
Finally it replaces all integrals by their tabulated reduction to a set of known master integrals~\cite{Czakon:2004bu}.
The reduction to master integrals is performed by \code{Crusher}~\cite{crusher} and relies on integration-by-parts identities relating integrals with different propagator powers through a system of coupled equations to each other~\cite{Chetyrkin:1981qh}.
The system of equations can be solved with the Laporta algorithm~\cite{Laporta:2001dd} such that all appearing integrals can be written in terms of a linear combination of a finite number of master integrals.

The final renormalization step during which all $Z_x$-factors -- with $x \in \{\phi, \phi^2, \phi^4, A, \phi A\phi\}$ --  are obtained is carried out by a \code{FORM} program,
which ensures the insertion of all possible counter terms in an automated way.

\renewcommand{\picscalefactor}{0.15}
\begin{table}[t]
\centering
\begin{tabular*}{\linewidth}{l@{\extracolsep\fill}rrrr}
  \hline\hline\\[-10pt]
Loops & 1 & 2 & 3 & 4 \\[4pt]
\hline\\[-12pt]
 \vcenteredhbox{\includegraphics[scale=\picscalefactor]{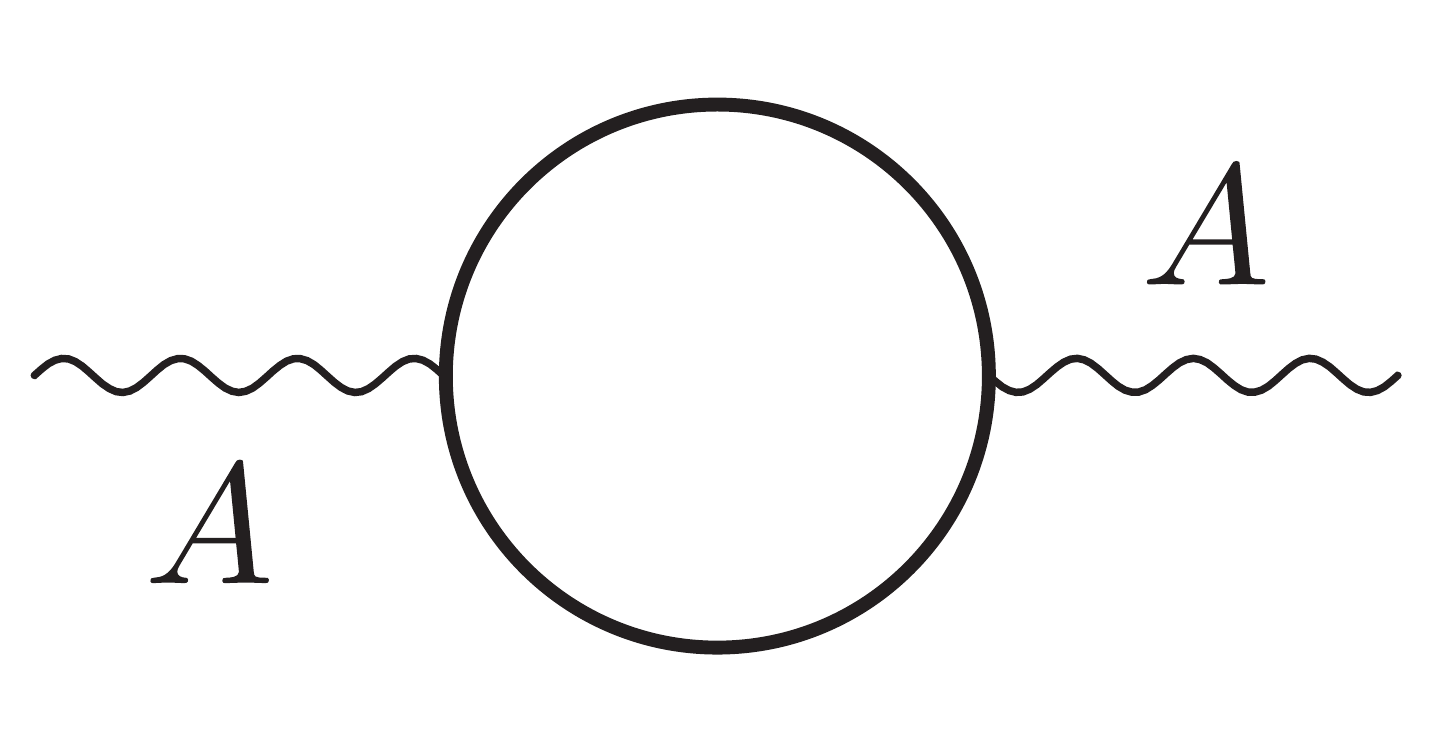}} & \vcenteredhbox{$2$} & \vcenteredhbox{$20$} & \vcenteredhbox{$370$} & \vcenteredhbox{$9,291$}\\[4pt]
\hline\\[-12pt]
 \vcenteredhbox{\includegraphics[scale=\picscalefactor]{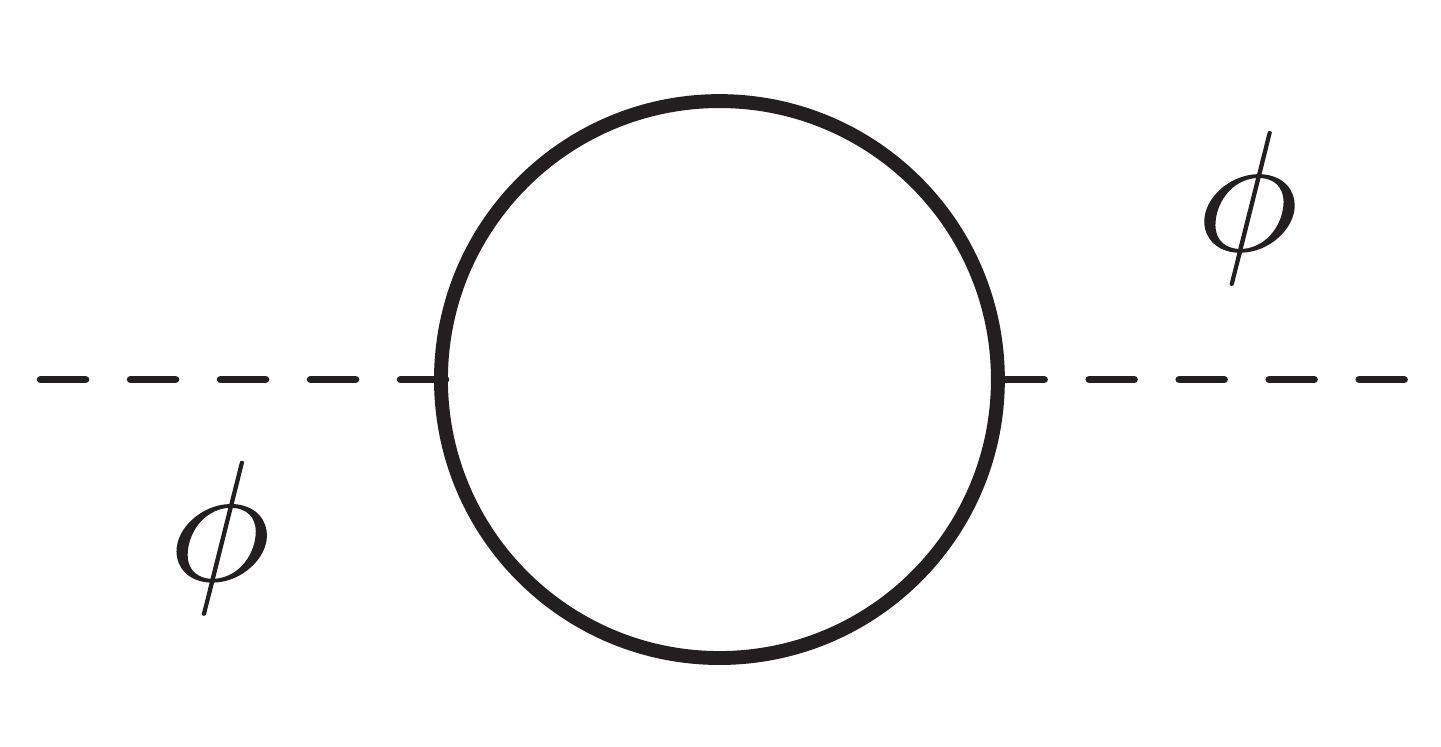}}  & \vcenteredhbox{$4$} & \vcenteredhbox{$27$} & \vcenteredhbox{$459$}& \vcenteredhbox{$11,332$}\\[4pt]
\hline\\[-12pt]
 \vcenteredhbox{\includegraphics[scale=\picscalefactor]{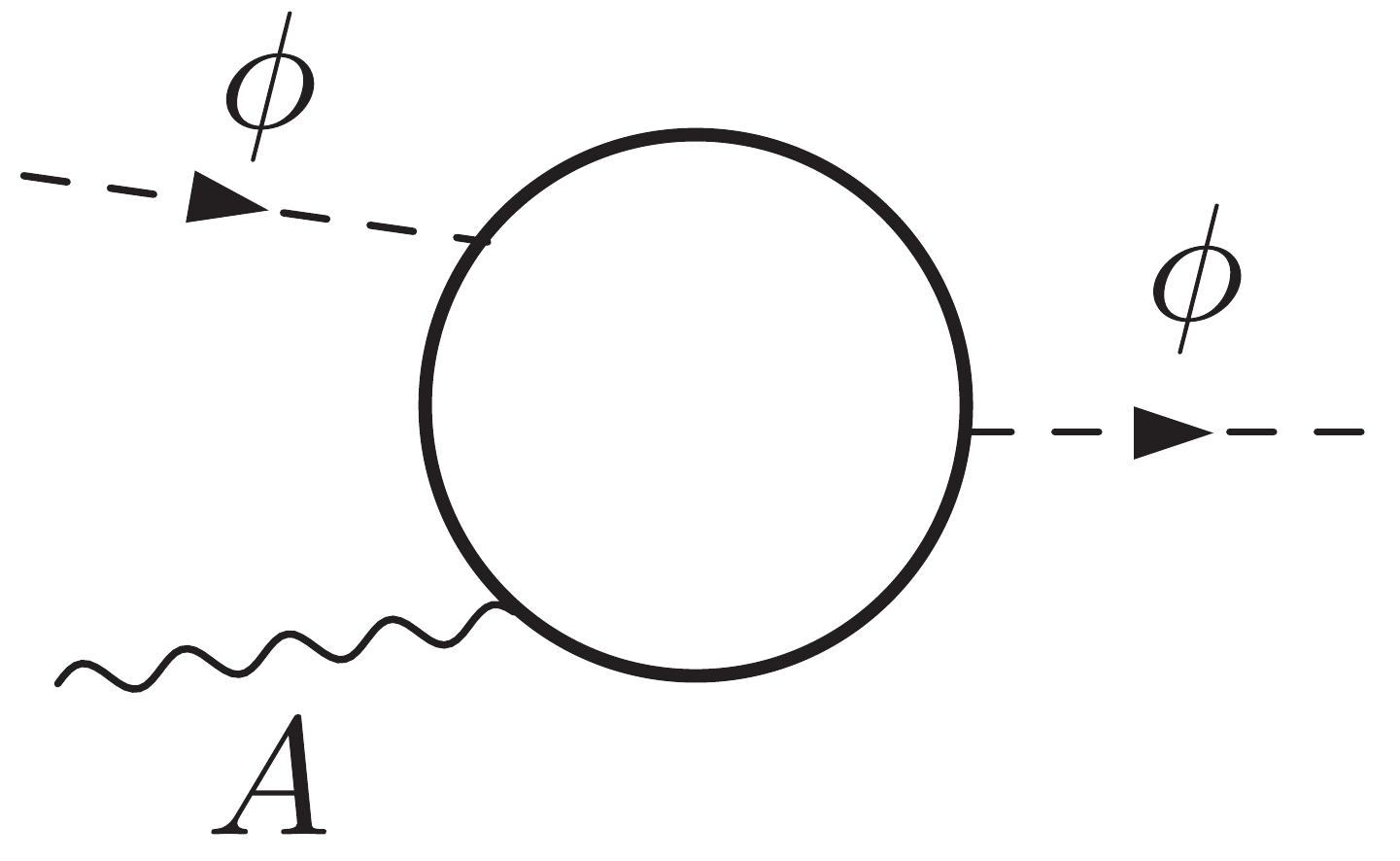}} & \vcenteredhbox{$5$} & \vcenteredhbox{$107$} & \vcenteredhbox{$3,078$}& \vcenteredhbox{$106,501$}\\[4pt]
\hline\\[-12pt]
 \vcenteredhbox{\includegraphics[scale=\picscalefactor]{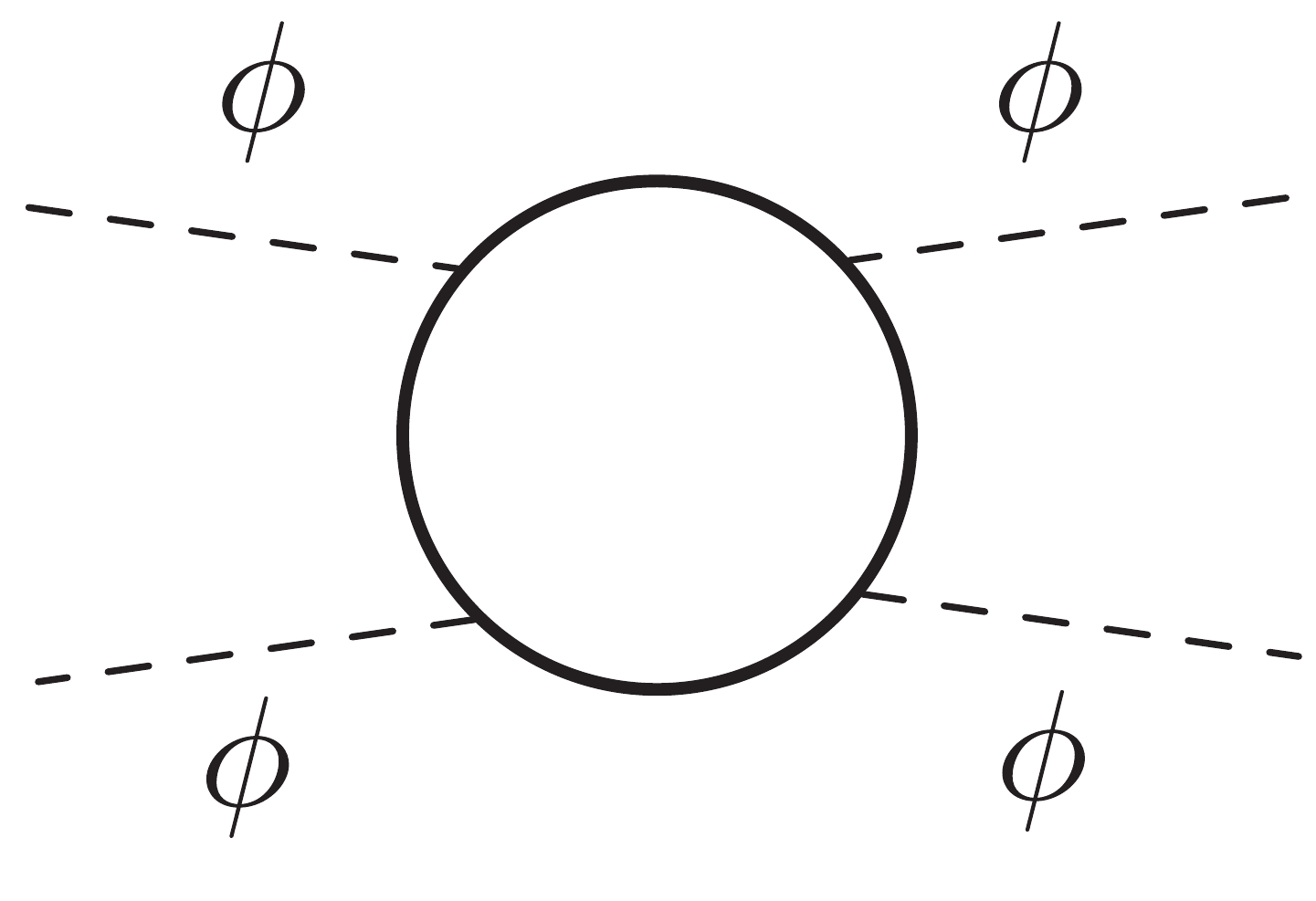}} & \vcenteredhbox{$32$} & \vcenteredhbox{$1,042$} & \vcenteredhbox{$40,164$}&  \vcenteredhbox{$1,735,706$}\\[4pt]
 \hline\hline
\end{tabular*}
 \caption{List of all relevant $n$-point functions and the associated number of Feynman diagrams evaluated in dependence on the number of loops ($A$: gauge field, $\phi$: scalar field).\label{TAB:DiagramListing}}
\end{table}

We explicitly computed the vertex renormalization constant $Z_{\phi A \phi }$ in order to check that the Ward identity $Z_{\phi}/Z_{\phi A \phi}=1$ holds
and the charge renormalization is solely determined by the wave-function renormalization constant $Z_A$ of the photon.
The mass square vertex renormalization  $Z_{\phi^2}$ of the scalar was obtained via a single zero momentum mass square operator insertion into the two-point function of the scalar.

The IRR procedure requires to keep track of the degree of the overall UV divergence via a global divergence power counting variable.
Besides the physical renormalization constants already mentioned, one has to insert artificial mass counter terms for the photon and scalar starting at the two-loop level.
At the one loop level the IRR introduces artificial divergent mass terms proportional to the unphysical regulator mass square, which can be easily dropped.
However, at the two-loop level such terms do get multiplied with negative powers of the regulator mass and thus lose their ``mass label'' and mix with physical divergences.
In order to prevent such a mixing one has to subtract these terms via explicit mass counter term insertions.
As a welcome check we verified that on the removal of the divergence power counting variable, each of the two artificial mass counter terms adds up to zero, order by order in the loop expansion.

We also kept the full dependence on the gauge parameter $\xi_P=1-\xi$ appearing in the photon propagator.
As welcome check we could verify that $\xi_P$ fully cancels in all beta function and anomalous dimensions except for $\gamma_{\phi},\gamma_{\phi^2},\gamma_{\phi^4}$ and $\gamma_{\phi A \phi }$,
where it is in principle known that $\xi_P$ appears only once at the one loop order~\cite{Kissler:2016tne}.

We further could verify that upon adopting the $\hat{G}$ scheme,
which involves the redefinition of odd zeta values in terms of an epsilon expansion
containing even zeta values -- in our case just $\hat\zeta_3=\zeta_3-\tfrac{3}{4}\zeta_4\epsilon$ -- all $\zeta_4$ terms vanished in our results for the $Z$-factors.
So all $\zeta_4$ terms in the $Z$-factors can be fully reconstructed from the existing $\zeta_3$ ones following the ``no-$\pi$ theorem''~\cite{Baikov:2018wgs,Kotikov:2019bqo}.

\section{Four-loop renormalization constants} \label{sec:beyondoneloop}

The beta functions up to three loops are given in Sec.~\ref{sec:rg}. Here, we provide the missing terms at four loops and give the full expressions for the anomalous dimensions and the flow of the gauge fixing parameter. The four-loop contributions to the beta functions read
\begin{widetext}
\begin{align}
   &\beta_\alpha^{(4\ell)} = \textstyle \left(-\frac{323 }{3888}n^3+\left[\frac{451}{54}-\frac{38 \zeta _3}{9}\right] n^2+\frac{3 }{16}n\right)\alpha ^5 +\left(\frac{5 }{72}n^3-\frac{41 }{9}n^2-\frac{37}{8} n\right)\alpha ^4 \lambda  +\left(\frac{1}{24}n^3+\frac{139 }{48}n^2+\frac{137 }{48}n\right)\alpha ^3 \lambda ^2\\
   &\hspace{1cm}\textstyle+\left(-\frac{5 }{48}n^3-\frac{7 }{16}n^2-\frac{1}{3}n\right)\alpha ^2 \lambda ^3 \nonumber \\[0.2cm]
   &\beta_\lambda^{(4\ell)} =\textstyle \left(\!\left[40 \zeta _5-\frac{27 \zeta _3}{2}-\frac{6145}{48}-\frac{\pi ^4}{180}\right] n^2 \!-\!\left[\frac{3157 \zeta _3}{6}+80 \zeta _5+\frac{25123}{24}\!+\!\frac{143 \pi ^4}{180}\right]\!n\!-\frac{12 \pi ^4}{5}-\frac{18503}{16}-768 \zeta _3-960 \zeta _5\right)\alpha ^2 \lambda ^3  \\
   & \textstyle + \left(\!\left[\frac{29 \zeta _3}{2}-\frac{\pi ^4}{60}-\frac{377}{96}\right] n^2+\left[-\frac{269 \zeta _3}{2}+50 \zeta _5+\frac{7 \pi ^4}{10}-\frac{1403}{32}\right] n+\frac{103 \pi ^4}{60}-\frac{185}{12}+150 \zeta _5-435 \zeta _3\right)\alpha  \lambda ^4 \nonumber \\
   & \textstyle +\left(\!\left[\frac{1}{81}-\frac{2 \zeta _3}{9}\right]\! n^3\!+\!\left[28 \zeta _3-\frac{\pi ^4}{10}-\frac{55709}{1296}\right] n^2+\left[310 \zeta _5-\frac{578 \zeta _3}{3}-\frac{13987}{36}-\frac{19 \pi ^4}{60}\right] n-\frac{12751}{16}-\frac{33 \pi ^4}{20}+504 \zeta _3-390 \zeta _5\right)\alpha ^5 \nonumber \\
   & \textstyle +\left(\!\left[\frac{19 \zeta _3}{18}\!+\!\frac{67}{2592}\!\right]\! n^3\!+\!\left[20 \zeta _5\!-\!\frac{77 \zeta _3}{2}\!+\!\frac{\pi ^4}{5}\!+\!\frac{12779}{162}\right]\! n^2\!+\!\left[\frac{431 \zeta _3}{2}\!+\!305 \zeta _5\!+\!\frac{13 \pi ^4}{10}\!+\!\frac{209}{12}\right]\! n\!+\!\frac{26 \pi ^4}{5}\!-\!\frac{19127}{96}\!+\!2095 \zeta _5-\frac{1191 \zeta _3}{2}\right)\alpha ^4 \lambda  \nonumber \\
   & \textstyle -\left(\!\left[\frac{7 \zeta _3}{6}\!+\!\frac{139}{1944}\!\right]\! n^3\!-\!\left[40 \zeta _5\!-\!\frac{157 \zeta _3}{3}-\frac{\pi ^4}{60}+\frac{289817}{7776}\!\right]\! n^2\!-\!\left[\!1080 \zeta _5\!-\!\frac{1813 \zeta _3}{2}\!-\!\frac{5 \pi ^4}{12}\!+\!\frac{88871}{96}\!\right]\! n\!+\!\frac{6 \pi ^4}{5}\!-\!\frac{66851}{48}\!+\!725 \zeta _3\!-\!1520 \zeta _5\!\right)\!\alpha ^3 \lambda ^2 \nonumber \\
   & \textstyle +\left(-\frac{5 }{96}n^3+\left[\frac{63 \zeta _3}{2}+20 \zeta _5-\frac{\pi ^4}{12}+\frac{395}{12}\right] n^2+\left[191 \zeta _3+275 \zeta _5+\frac{10057}{48}-\frac{31 \pi ^4}{60}\right] n-\frac{11 \pi ^4}{15}+\frac{24581}{96}+\frac{583 \zeta _3}{2}+465 \zeta _5\right)\lambda ^5  \nonumber
\end{align}
\end{widetext}
The field anomalous dimensions are defined through the relation
$\gamma_x = \frac{d\ln Z_x}{d\ln b}$,
for $x \in \{\phi,\phi^2,A_\mu\}$. Schematically, order by order, they can be written as
$\gamma_x = \sum_{i = 1}^L \gamma_{x}^{(i\ell)}$.
Up to three loops, the contributions to $\gamma_{\phi^2}$ read
\begin{align}
  \gamma_{\phi^2}^{(1\ell)} & = (1-\xi ) \alpha  -  (n+1) \lambda \,,\\
   \gamma_{\phi^2}^{(2\ell)} & =  \frac{3}{2}(n+1)  \lambda ^2-\!\left(\!5 n+\frac{1}{2}\right)\alpha ^2 -4 (n+1) \alpha  \lambda  \,,\\
    \gamma_{\phi^2}^{(3\ell)} & = \left(6 \zeta _3 (3 n+2)-\frac{167 n}{4}-\frac{51}{2}\right)\alpha ^3 \\
      &+\frac{1}{16}  (n+1) \left(144 \zeta _3+43 n+299\right)\alpha ^2 \lambda \nonumber \\
      &+3   \left(\zeta _3+1\right)(n+1)\alpha \lambda ^2 -\!\frac{(n+1)}{16}   (31 n+115)\lambda ^3\,. \nonumber
\end{align}
The contributions to $\gamma_{\phi}$ are
\begin{align}
  \gamma_{\phi}^{(1\ell)} & = -  (\xi +2)\alpha\,,
 \end{align}
 \begin{align}
   \gamma_{\phi}^{(2\ell)} & = \frac{1}{12} (11 n+9) \alpha ^2+\frac{1}{4}  (n+1)\lambda ^2\,, \\[5pt]
    \gamma_{\phi}^{(3\ell)} & = \left(\frac{1}{432} n (5 n+3267)+\frac{1}{8}-3 \zeta _3 (n-1)\right)\alpha ^3 \nonumber\\
     &-\frac{1}{8} \left(24 \zeta _3-13\right) (n+1) \alpha ^2\lambda  +\frac{5}{4}  (n+1)\alpha  \lambda ^2 \nonumber \\
     &-\frac{1}{16}  (n+1) (n+4)\lambda ^3\,,
\end{align}
and for $\gamma_{A}$ we obtain
\begin{align}
  \gamma_{A}^{(1\ell)}&=\frac{ n}{3}\alpha \,,\quad   \gamma_{A}^{(2\ell)} = 2 n \alpha^2\,,\\[5pt]
    \gamma_{A}^{(3\ell)}&=\frac{n }{72}  (261-49 n)\alpha ^3+\frac{n}{2}(n+1)\alpha ^2 \lambda\notag\\
    &-\frac{n}{8}(n+1)\alpha  \lambda ^2\,.
\end{align}
The contributions at four loops to these field anomalous dimensions are, again, more lengthy.  They read
%
\begin{widetext}
\begin{align}
  &\gamma_{\phi^2}^{(4\ell)} = \textstyle  \left(\left[-\frac{127 \zeta _3}{6}+40 \zeta _5-\frac{1889}{36}-\frac{\pi ^4}{45}\right] n^2+\left[-\frac{368 \zeta _3}{3}+70 \zeta _5-\frac{5249}{36}-\frac{67 \pi ^4}{180}\right] n-\frac{280}{3}-\frac{7 \pi ^4}{20}+30 \zeta _5-\frac{203 \zeta _3}{2}\right)\alpha ^2 \lambda ^2 \notag\\
  &\textstyle + \left(-\frac{55 \zeta _3}{2}+\left[-\frac{9 \zeta _3}{2}+\frac{\pi ^4}{30}-\frac{775}{96}\right] n^2+\left[-32 \zeta _3+\frac{\pi ^4}{12}-\frac{369}{16}\right] n+\frac{\pi ^4}{20}-\frac{1439}{96}\right) \alpha  \lambda ^3 \nonumber \\
  &\textstyle+ \left(\frac{49 \zeta _3}{4}+70 \zeta _5+\left[\frac{\zeta _3}{2}+\frac{5}{324}\right] n^3+\left[-\frac{145 \zeta _3}{6}+\frac{\pi ^4}{8}+\frac{42545}{1296}\right] n^2+\left[77 \zeta _3-\frac{285 \zeta _5}{4}+\frac{23 \pi ^4}{60}+\frac{97}{24}\right] n+\frac{11 \pi ^4}{40}-\frac{1283}{48}\right)\alpha ^4 \nonumber \\
  &\textstyle+  \left(120 \zeta _5-\frac{327 \zeta _3}{2}-\left[\frac{7 \zeta _3}{6}+\frac{139}{1944}\right]\! n^3\!+\!\left[\frac{273683}{7776}-\frac{119 \zeta _3}{6}-\frac{\pi ^4}{30}\right]\! n^2\!+\!\left[120 \zeta _5+\frac{\pi ^4}{15}+\frac{23665}{144}-\frac{1093 \zeta _3}{6}\right]\! n\!+\!\frac{\pi ^4}{10}+\frac{12391}{96}\right) \alpha ^3 \lambda \nonumber \\
  &\textstyle +\left(\frac{17 \zeta _3}{4}+\left[\frac{3 \zeta _3}{4}-\frac{1}{12}\right] n^3+\left[2 \zeta _3+\frac{\pi ^4}{24}+\frac{977}{48}\right] n^2+\left[\frac{11 \zeta _3}{2}+\frac{2 \pi ^4}{15}+\frac{377}{6}\right] n+\frac{11 \pi ^4}{120}+\frac{2035}{48}\right) \lambda ^4\,, \\[0.5cm]
  &\gamma_\phi^{(4\ell)} = \textstyle \left(-4 \zeta _3-20 \zeta _5+\left[\frac{5 \zeta _3}{3}+\frac{199}{144}-\frac{\pi ^4}{180}\right] n^2+\left[-\frac{7 \zeta _3}{3}-20 \zeta _5+\frac{2 \pi ^4}{45}+\frac{413}{18}\right] n+\frac{\pi ^4}{20}+\frac{345}{16}\right) \alpha ^3 \lambda   \notag\\
  &\textstyle + \left(\frac{11 \zeta _3}{2}+5 \zeta _5+\left[\frac{19 \zeta _3}{12}-\frac{\pi ^4}{120}-\frac{641}{288}\right] n^2+\left[\frac{85 \zeta _3}{12}+5 \zeta _5-\frac{179}{18}-\frac{\pi ^4}{24}\right] n-\frac{\pi ^4}{30}-\frac{247}{32}\right)\alpha ^2 \lambda ^2 \nonumber \\
  &\textstyle + \left(-\zeta _3+\left[\frac{\zeta _3}{2}-\frac{19}{96}\right] n^2+\left[\frac{61}{96}
  -\frac{\zeta _3}{2}\right] n+\frac{5}{6}\right)\alpha  \lambda ^3 \nonumber + \left(-\frac{5 }{64}n^3
  +\frac{5 }{8}n^2+\frac{85 }{32}n+\frac{125}{64}\right)\lambda ^4 \\
  &\textstyle +\left(\frac{63 \zeta _3}{2}-45 \zeta _5+\left[\frac{13}{5184}-\frac{\zeta _3}{36}\right] n^3+\left[\frac{9 \zeta _3}{4}
  -\frac{\pi ^4}{120}-\frac{1505}{432}\right] n^2+\left[-\frac{49 \zeta _3}{4}-\frac{25 \zeta _5}{4}+\frac{231}{16}-\frac{\pi ^4}{24}\right] n
  -\frac{\pi ^4}{20}+\frac{133}{64}\right)\alpha ^4\,, \\[0.5cm]
  &
  \gamma_A^{(4\ell)} = \textstyle \left(\frac{323 }{3888}n^3+\left[\frac{38 \zeta _3}{9}-\frac{451}{54}\right] n^2-\frac{3 }{16}n\right)\alpha ^4
  + \left(-\frac{5 }{72}n^3+\frac{41 }{9}n^2+\frac{37 }{8}n\right) \alpha ^3 \lambda  + \left(-\frac{1}{24}n^3-\frac{139 }{48}n^2
  -\frac{137 }{48}n\right)\alpha ^2 \lambda ^2 \notag\\
  &\textstyle + \left(\frac{5 }{48}n^3+\frac{7 }{16}n^2+\frac{1}{3}n\right)\alpha  \lambda ^3\,.
\end{align}
\end{widetext}
%
Finally, we also give the corresponding expression for the renormalization group flow of the gauge fixing parameter
\begin{align}
  \beta_\xi^{(1\ell)} &= \frac{8}{3} n \alpha   \xi\,,\\
  \beta_\xi^{(2\ell)} &= 16  n \alpha ^2  \xi\,, \\
  \beta_\xi^{(3\ell)} &= \left(29 -\frac{49 n}{9}\right) n\alpha ^3\xi\\
  &\quad+\left(\frac{17}{6} \alpha - \frac{5}{12}  \lambda \right) n (n+1) \alpha \lambda\, \xi\,, \notag\\
   \beta_\xi^{(4\ell)} &= \xi\Big(\frac{11}{192} n \left(5 n^2+21 n+16\right) \alpha  \lambda ^3\\
   &-\frac{25}{288} n (5 n^2 -328 n-333 )\alpha ^3 \lambda\notag\\
   & -\frac{3}{32}  n \left(2 n^2+139 n+137\right)\alpha ^2 \lambda ^2 \nonumber \\
  & +\frac{n}{486} \left(323 n^2+72 n (228 \zeta_3-451)-729\right)\!\alpha ^4\Big)\,. \nonumber
\end{align}

\section{Pad\'e and Pad\'e-Borel approximants}\label{app:pade}

We explore various resummations for the series
\begin{align}\label{eq:ncinf}
  n_c =  n_{c,0}+ n_{c,1}\epsilon + n_{c,2}\epsilon^2 + n_{c,3}\epsilon^3+ \mathcal{O}(\epsilon^4)\,,
\end{align}
from Eq.~\eqref{eq:ncseries}. The Pad\'e approximants are defined as
\begin{align}
  [m/n] = \frac{a_0 + a_1 \epsilon + \dots + a_m \epsilon^m}{1 + b_1 \epsilon + \dots + b_n \epsilon^n}\,,
\end{align}
with $L -1= m+n$.
The real-valued coefficients $\{a_i, b_i\}$ are uniquely determined by matching its Taylor series to the original expansion, e.g., Eq.~\eqref{eq:ncinf}, order by order.

Assuming the series is Borel summable, we also compute the corresponding  Pad\'e-Borel approximants.
The Borel transform is defined as
\begin{align}
  B_{n_c}(\epsilon) = \sum\limits_{k=0}^3 \frac{n_{c,k}}{k!} \epsilon^k\,. \label{eq: Borel transform}
\end{align}
Using $B_{n_c}$, one analytically continues to the real axis by calculating the Borel sum
\begin{align}
  \tilde{n}_c(\epsilon) := \int_0^\infty \mathrm{d}t \, \mathrm{e}^{-t} B_{n_c}(\epsilon t) \,. \label{eq: Borel sum}
\end{align}
Approximating the Borel transform by a Pad\'e approximant, the Pad\'e-Borel resummed series is obtained.

\begin{figure}[t!]
  \includegraphics{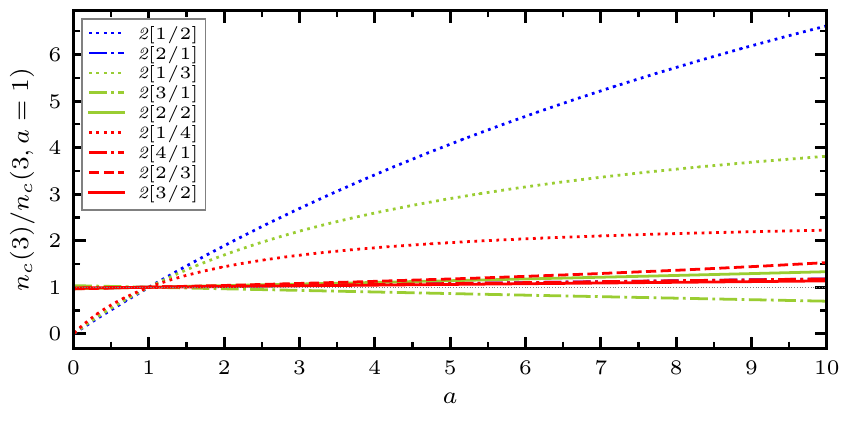}
  \caption{Dependence of  the two-sided Pad\'e approximants  on the slope $a$ near the lower critical dimension evaluated at $d=3$. All values are normalized to their corresponding estimate at $a=1$ as presented in Tab.~\ref{tab: interpolations}. See also Tab.~\ref{tab: a dep twosided Pade} for a list of
  estimates of the highest order two-sided Pad\'es (red) at different $a$.}
  \label{fig: a dep twosided Pade}
\end{figure}

For the interpolation between the epsilon expansions at lower and upper critical dimension we construct the two-sided Pad\'e approximants. To that end, the plain Pad\'e approximants are generalized making the ansatz
\begin{align}
      \textit{2}_{[m/n]}(d) = \frac{a_0 + a_1 d + \dots + a_m d^m}{1+ b_1 d + \dots + b_n d^n}\,.
\end{align}
Herein, the coefficients $\{a_i\}$ and $\{b_j \}$ are fixed such that the function matches the epsilon expansions near $d=2$ and $d=4$. This leads to the relations
\begin{align}
      \textit{2}_{[m/n]}^{(1)}(2)= a \,,\quad \textit{2}_{[m/n]}^{(k)}(4)= (-1)^k k! n_{c,k}\,,
\end{align}
with $n_{c,k}$ from Eq.~\eqref{eq:ncinf} and, again, the variable $a>0$ describes the unkown slope of $n_c$ close to $d=2$.
In the text we provide estimates for $n_c$ at $d=3$ by setting $a=1$ and in Fig.~\ref{fig: a dep twosided Pade} we show the $a$ dependence of two-sided Pad\'e evaluated at $d=3$. Additionally, we compiled
the estimates of the highest order two-sided Pad\'es for different $a$ in Tab.~\ref{tab: a dep twosided Pade}.
\begin{table}[t!]
  \caption{Two-sided Pad\'e estimates from four loops at $d=3$ and different slopes $a$ for the expansion in $d=2+\epsilon$. In the last
  column we show the mean value of the two-sided Pad\'es at a specific $a$ and its standard deviation.
  Slanted numbers denote non-monotonic curves between $d\in(2,4)$.}
  \label{tab: a dep twosided Pade}
  \begin{tabular*}{\linewidth}{r@{\extracolsep\fill}rrrrr}
    \hline\hline
    $a$ & $\textit{2}[2/3]$ & $\textit{2}[3/2]$ & $\textit{2}[1/4]$ & $\textit{2}[4/1]$ & $\approx$ \\
    \hline
    0.2 & 15.85 & 13.93 & 2.84 & \textit{7.93} & 10.14 $\pm$ 5.92\\
    0.5 & 16.02 & 14.00 & 6.08 & \textit{7.98} & 11.02 $\pm$ 4.75\\
    1   & 16.32 & 14.11 & 9.80 & \textit{8.06} & 12.07 $\pm$ 3.80\\
    2   & 16.94 & 14.32 & 14.12 & \textit{8.22} & 13.40 $\pm$ 3.68\\
    5   & 19.21 & 14.97 & 19.20 & \textit{8.71} & 15.52 $\pm$ 4.96\\
    10  & 24.98 & 16.03 & 21.82 & \textit{9.51} & 18.08 $\pm$ 6.81\\
    \hline\hline
  \end{tabular*}
\end{table}

%

\section{Borel resummation algorithm}\label{eq:borel}

Borel resummation with conformal mapping~\cite{zinn1996quantum} facilitates high-accuracy determination of critical exponents in 3$d$ $O(n)$ symmetric $\phi^4$ theories from the
$(4-\epsilon)$ expansion~\cite{Kompaniets2017}.
For the present application, we make the assumption that the AH model's asymptotic behavior is qualitatively the same as the one from the ungauged scalar models. More explicitly,we assume that the epsilon expansion for the critical number of components follows a formal power series with factorially increasing coefficients, i.e.
\begin{align}
      n_{c,k} \sim (-a)^k \mathrm{\Gamma}(k+b+1) \approx  (-a)^k k! k^b \label{eq:largeordercoeff}
\end{align}
for large $k$. The parameter $a$ was determined for the AH model to
$a \approx 0.0808$~\cite{Itzykson1977} .
The generalized Borel(-Leroy) transform \cite{Kleinert} of this series  is calculated as in Eq.~\eqref{eq: Borel transform} by canceling the factorial growth of the coefficients
\begin{align}
      \mathcal{B}^b_{n_c}(\epsilon) := \sum\limits_{k=0}^\infty \frac{n_{c,k}}{\mathrm{\Gamma}(k+b+1)}\epsilon^k = \sum\limits_{k=0}^\infty B_k^b \epsilon^k\,.
      \label{eq: BorelLeroy transform}
\end{align}
The coefficients behave like $B_k^b \sim (-a)^k$ and therefore follow a geometric series which can written as rational function with a pole at $\epsilon=-1/a$
\begin{align}
      \mathcal{B}^b_{n_c}(\epsilon) \underset{\text{$k$ large}}{\sim} \sum\limits_{k} (-a)^k \epsilon^k = \frac{1}{1+a \epsilon}\,.
\end{align}
This increases the originally vanishing radius of convergence to in a circle in the complex plane with $|\epsilon|<1/a$.
The assumption of Borel summability now implies that the Borel transform can be analytically continued to the positive real axis such that the Borel sum \eqref{eq: Borel sum} -- with an extra factor $t^b$ in the integrand --
is convergent and leads to the correct estimate for $\epsilon=1$.

In order to incorporate the large-order behavior of the asymptotic series, in the next step the Borel transform is analytically continued with a conformal transformation to the real axis, i.e.
\begin{align}
  w(\epsilon) = \frac{\sqrt{1+a \epsilon} -1}{\sqrt{1+ a \epsilon} +1} \quad \Rightarrow \quad \epsilon = \frac{4}{a}\frac{w}{(1-w)^2}\,.
  \label{eq: conf mapping}
\end{align}
This maps the cut of the negative real axis $(-\infty, -1/a)$ of the Borel transform to the unit circle in the variable $w$.
Expanding the Borel transform in $w$ provides a convergent series in the  integration domain of the Borel sum.

As a further improvment for finite order truncated Borel sums, we introduce the parameter $\lambda$~\cite{Kompaniets2017,Vladimirov1979},
\begin{align}
      \mathcal{B}^b_{n_c}(\epsilon) \approx \mathcal{B}^{a,b,\lambda}_{n_c}(\epsilon) := \left( \frac{a \epsilon(w)}{w} \right)^\lambda \sum\limits_{k=0}^L B_{k}^{b,\lambda} w^k\,.
      \label{eq:Borelsumlambda}
\end{align}
which restores the actual large $\epsilon$ behavior $\sim\epsilon^\lambda$ of the Borel transform \cite{Kompaniets2017}.
In the analysis presented here, the parameter $\lambda$ is used to improve the sensitivity of the resummation algorithm, see below.
On top of that, a homographic transformation with shifting variable  $q$
\begin{align}
      \epsilon = h_q(\tilde{\epsilon}) := \frac{\tilde{\epsilon}}{1+q \tilde{\epsilon}} \quad \Rightarrow \quad \tilde{\epsilon} = h^{-1}_q(\epsilon) = \frac{\epsilon}{1-q \epsilon}\,,
\end{align}
can be used to optimize the numerical stability. The resummed series is therefore determined by
\begin{align}
      \tilde{n}_c(\epsilon) \approx {n_c}_L^{a,b,\lambda,q}(\epsilon) := \! \int\limits_{0}^\infty \! \mathrm{d}t \, t^b \mathrm{e}^{-t} \mathcal{B}^{a,b,\lambda}_{n_{c}\circ h_q}(h^{-1}_q(\epsilon) t)\,.\notag
\end{align}
%
%
The set of resummation parameters $a$, $b$, $\lambda$ and $q$ are optimized by measuring the sensitivity of the summation to a parameter change.
For a function $F$ the variation upon changing  one of its parameters $x$ in a range $x\in[x_0, x_0+\Delta]$ can be defined as
\begin{align}
      \mathcal{S}_x(F(x_0)) := \min\limits_{x\in[x_0,x_0+\Delta]}\left( \max\limits_{x'\in[x_0,x_0+\Delta]} \left| F(x) - F(x') \right| \right)\,.\notag
\end{align}
For a function $F(x)$, which is weakly varying within a plateau of length $\Delta$, $\mathcal{S}_x$ is small.
Therefore, we can quantify the sensitivity as the extension of the plateaus. At the highest available order $\mathcal{O}(\epsilon^3)$, we find the  spreads
\begin{align}
       \Delta_b \approx 30,  \quad        \Delta_\lambda \approx 0.3,     \quad     \Delta_q \approx 0.02\,. \label{eq: plateau length}
\end{align}
The optimal set of parameters is identified by making use of both, the ``principle of minimal sensitity'' and ``principle of fastest convergence''~\cite{Kompaniets2017}.
\begin{figure}[t!]
  \includegraphics[width=0.48\columnwidth]{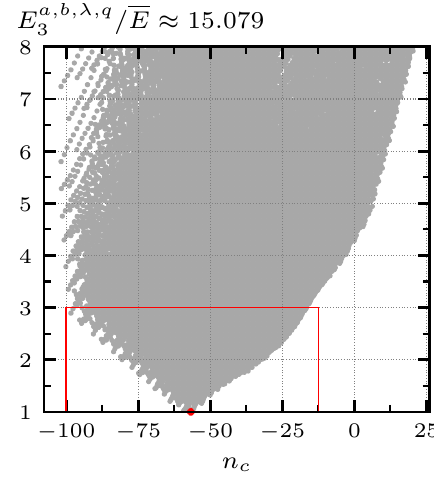}\hfill
  \includegraphics[width=0.48\columnwidth]{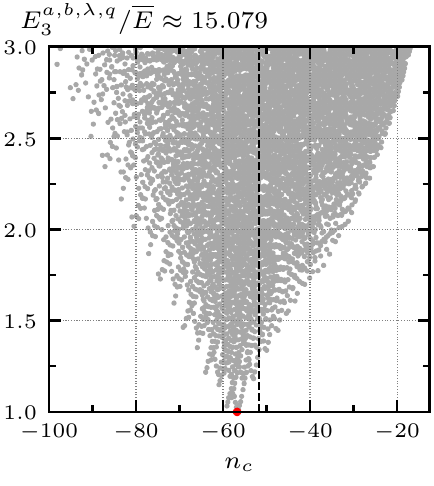}
  \caption{Borel scan for the critical number of components. The parameters $b$,$\lambda$ and $q$ are varied in the interval
  $b \times \lambda \times q \in [0,90] \times [0,4] \times [0,0.4]$ while $a$ was fixed to $0.0808$ \cite{Itzykson1977}. For each point in the spanned parameter space the error estimate was computed. The parameter set with the smallest error lies at $\overline{E}(87,0.66,0.0) \approx 15.079$ and yield $n_c\approx-57$. For points below a relative error of $E_i/\overline{E} < 3$ (right panel) we compute the
  weighted mean to $\tilde{n}_c \approx-52$ (black dotted line).}
  \label{fig: Borel scan}
\end{figure}
The different variations are collected in an error estimate~\cite{Kompaniets2017,Ihrig2018},
\begin{align}
      E_{{n_c},L}^{b,\lambda,q} := &\max\left\{ |{n_c}_{L}^{b,\lambda,q}-{n_c}_{L-1}^{b,\lambda,q}|, |{n_c}_{L}^{b,\lambda,q}-{n_c}_{L-2}^{b,\lambda,q}|  \right\} \nonumber \\
            &+ \max\left\{ \mathcal{S}_b\left({n_c}_L^{b,\lambda,q}\right), \mathcal{S}_b\left({n_c}_{L-1}^{b,\lambda,q}\right) \right\} \nonumber \\
            &+ \mathcal{S}_\lambda\left( {n_c}_L^{b,\lambda,q}\right) +\mathcal{S}_q\left( {n_c}_L^{b,\lambda,q}\right) \,.
            \label{eq: error estimate}
\end{align}
whichs is minimized by the wanted set of parameters. This includes als the sensitivities at different orders $L$ and  search
for a minimum in the dependence on $L$.

Explicitly, we scan the parameter space in the range
$(b,\lambda,q) \in [0,90] \times [0,4] \times [0, 0.4]$
in steps of $\delta_b = 1$, $\delta_\lambda = 0.02, \delta_q = 0.02$ and compute $E_{{n_c},L}^{b,\lambda,q}$ for each parameter set, see Fig.~\ref{fig: Borel scan}.
The gobal minimum $\overline{E}\approx 15.079$ of this scan marks the apparently best set of resummation parameters and
is found at $(b,\lambda,q)~=~(87.0.66,0.0)$.
However, this minimum is not always sharp and there are other sets of parameters which are almost equally likely.
Therefore, we compute a weighted mean of all points which lie below a relative error of $E_i/\overline{E}<3$ as in Ref.~\cite{Ihrig2018}.
From the error estimate at the minimum we compile the error for the resummation of $n_c$ at $d=3$ with error as $3\overline{E}$ to
$\tilde{n}_c (d=3) \approx -52 \pm 45$.

\section{Relation to the $\mathbb{CP}^{n-1}$ model }\label{app:cp1}

The AH model is related to the $\mathbb{CP}^{n-1}$ model with action
$S=\frac{1}{2t}\int \mathrm{d}^d x \left( \partial_\mu \bar{\phi}^a_i \partial_\mu \phi^a_i
        + (\bar{\phi}^a_i \partial_\mu \phi^a_i) (\bar{\phi}^b_i \partial_\mu \phi^b_i)\right)$
and constraint $\sum_{a=1}^n \bar{\phi}^a_i \phi^a_j =\delta_{ij}$~\cite{Hikami1979,March-Russell1992}.
The relation can be established by considering the non-linear sigma model (NL$\sigma$M) defined on the Grassmannian manifold of $U(n)/[U(n-p) \times U(p)]$. For $p=1$, the symmetric space is isomorphic to
$\mathbb{CP}^{n-1}$~\cite{Hikami1981}.
The beta function $\beta_{t}$ for general $p$ was computed in $d=2+\epsilon$ to four-loop order~\cite{Hikami1981, Hikami1983, Wegner1989}.
%
%
It features an IR-unstable fixed point $\beta_{t}(t_c) =0$ reading
\begin{align}
  t_c\!=\!\frac{\epsilon}{n}\!-\!\frac{2\epsilon^2}{n}\!+\!\frac{3(n-4)}{2n^3}\epsilon^3\!-\!\frac{\left(2 n^2-51 n+126\right)}{6 n^4}\epsilon ^4\!+\!\mathcal{O}(\epsilon^5),\notag
\end{align}
which is real valued for all $n > 0$. The associated correlation-length exponent $1/\nu$ is
\begin{align}
  \frac{1}{\nu}\! =\!\epsilon\!+\!\frac{2 }{n}\epsilon ^2\!+\!\frac{(3 n-4) }{n^2}\epsilon ^3\!+\!
    \left(\frac{11}{n^3}-\frac{9}{2 n^2}+\frac{1}{n}\right) \epsilon^4\!+\!\mathcal{O}\left(\epsilon ^5\right).\notag
\end{align}
A further connection can be drawn between the above NL$\sigma$M with target space $U(n)/[U(n-p)\times U(p)]$ and a $SU(p)\times U(1)$ gauge theory with Lagrangian~\cite{Hikami1980,March-Russell1992}
\begin{align}
  \hspace{-0.2cm}\mathcal{L}\!=\!|D_\mu \phi_a|^2\!+\!\textstyle \frac{F^2}{4}\!+\!\frac{G^2}{4}\!+\!\lambda (\bar{\phi}_a \phi_a)^2\!+\!\gamma (\bar{\phi}_a \phi_b) (\bar{\phi}_b \phi_a).\label{eq: SUp U1 gauge theory}
\end{align}
Here, $D_\mu = \partial_\mu - \mathrm{i}e A_\mu - \mathrm{i} g B_\mu$ is the covariant derivative and  $F^2$ ($G^2$)
are kinetic terms for the $U(1)$ ($SU(p)$) gauge fields $A_\mu$  ($B_\mu$).
Eq.~\eqref{eq: SUp U1 gauge theory} represents a generalization of the AH model and its beta functions can be deduced from it upon replacing
$n \rightarrow p n$ supplemented by the beta function for the $SU(p)$ gauge field~\cite{March-Russell1992}.
For $p=1$ we recover the AH model beta functions.
The large-$n$ expansion for Eq.~\eqref{eq: SUp U1 gauge theory} gives correlation-length exponent~\cite{Hikami1980}
\begin{align}
  \nu = \frac{1}{d-2}\left( 1 + \frac{2(d^2-d)\sin(d \pi/2)\Gamma(d-1)p}{\pi n \Gamma^2(d/2)} \right)\,, \notag
\end{align}
and anomalous dimension
\begin{align}
  \eta\!=\!\frac{p}{2n}\!\Big(\frac{2(d-4)\sin(\pi d/2) \Gamma(d-1)}{\pi d \, \Gamma^2(d/2)} \Big)\!\Big(\!1\!-\!\frac{4(d-1)^2}{4-d}\!\Big).\notag
\end{align}
As a check, we expanded the critical exponents of the AH model to first order in $1/n$ and find order by order agreement
in $\epsilon$.
Further, the large-$n$ inverse correlation-length exponent expanded around $d=2+\epsilon$  agrees with the
exponent from the $\mathbb{CP}^{n-1}$ model to all availabe orders in  $\epsilon$.  This corroborates the evidence that both theories lie in the same universality class for large $n$.

\bibliography{abelianhiggs}

\end{document}